\documentclass[aps,prx,superscriptaddress,twocolumn,showpacs]{revtex4-1}
\usepackage[table]{xcolor}
\usepackage{pgf}
\usepackage[utf8]{inputenc}
\usepackage{amsmath}
\usepackage{braket}
\usepackage{amssymb}
\usepackage{amsmath}
\usepackage{bbold}
\usepackage{enumitem} 
\usepackage{appendix}
\usepackage{graphicx}
\usepackage{overpic}
\usepackage{mathtools}
\usepackage{commath}
\usepackage{epstopdf}
\usepackage{cancel}
\usepackage[pdftex,bookmarks,colorlinks]{hyperref}
\usepackage{hyperref}
\usepackage{ulem}

\newcommand{\mean}[1]{\left\langle #1\right\rangle}
\newcommand{\Tr}{\operatorname{Tr}}
\newcommand{\nep}{\textrm{e}}
\newcommand{\ud}{\mathrm{d}}

\begin{document}

\title{Discrete truncated Wigner approach to dynamical phase transitions in Ising models after a quantum quench}

\author{Reyhaneh Khasseh}
\affiliation{Abdus Salam ICTP, Strada Costiera 11, I-34151 Trieste, Italy}
\affiliation{Department of Physics, Institute for Advanced Studies in Basic Sciences (IASBS), Zanjan 45137-66731, Iran}

\author{Angelo Russomanno}
\affiliation{Max-Planck-Institut f\"ur Physik Komplexer Systeme,N\"othnitzer Strasse 38, D-01187, Dresden, Germany}

\author{Markus Schmitt}
\affiliation{Department  of  Physics,  University  of  California  at  Berkeley,  Berkeley,  CA  94720,  USA}

\author{Markus Heyl}
\affiliation{Max-Planck-Institut f\"ur Physik Komplexer Systeme,N\"othnitzer Strasse 38, D-01187, Dresden, Germany}

\author{Rosario Fazio}
\affiliation{Abdus Salam ICTP, Strada Costiera 11, I-34151 Trieste, Italy}
\affiliation{Dipartimento di Fisica, Universit{\`a} di Napoli ``Federico II'', Monte S. Angelo, I-80126 Napoli, Italy}

\begin{abstract}
By means of the discrete truncated Wigner approximation we study dynamical phase transitions arising in the steady state of transverse-field Ising models after a quantum quench.
Starting from a fully polarized ferromagnetic initial condition these transitions separate a phase with nonvanishing magnetization along the ordering direction from a disordered symmetric phase upon increasing the transverse field.
We consider two paradigmatic cases, a one-dimensional long-range model with power-law interactions $\propto 1/r^{\alpha}$ decaying algebraically as a function of distance $r$ and a two-dimensional system with short-range nearest-neighbour interactions.
In the former case we identify dynamical phase transitions for $\alpha \lesssim 2$ and we extract the critical exponents from a data collapse of the steady state magnetization for up to 1200 lattice sites.
We find identical exponents for $\alpha \lesssim 0.5$, suggesting that the dynamical transitions in this regime fall into the same universality class as the nonergodic mean-field limit.
The two-dimensional Ising model is believed to be thermalizing, which we also confirm using exact diagonalization for small system sizes.
Thus, the dynamical transition is expected to correspond to the thermal phase transition, which is consistent with our data upon comparing to equilibrium quantum Monte-Carlo simulations.
We further test the accuracy of the discrete truncated Wigner approximation by comparing against numerically exact methods such as exact diagonalization, tensor network as well as artificial neural network states and we find good quantitative agreement on the accessible time scales. {Finally, our work provides an additional contribution to the understanding of the range and the limitations of
qualitative and quantitative applicability of the discrete truncated Wigner approximation.}
%
\end{abstract}

\maketitle

\section{Introduction}
Recent impressive developments underline the rich phase structures that can be generated by forcing isolated quantum matter out of equilibrium.
Some examples of these phenomena are the emergence of exotic phases, loss of adiabaticity across critical points in the context of the Kibble-Zurek 
mechanism and non-equilibrium phase transitions. These are some of the multiple aspects currently at the centre of an intense theoretical and experimental 
activity, as summarized in the reviews~\cite{silva_rmp,2008Bloch,abanin_rmp,Eisert2015,Heyl_2018,khemani2019brief,2019Altman}.

A paradigmatic protocol to drive a many-body system out of equilibrium, routinely used in experiments and intensively studied theoretically, is a quantum quench.
After initializing the system in a state, that can be thought of as a ground state of a given initial Hamiltonian, it is let evolving after an abrupt change of a Hamiltonian 
parameter. The long-time steady states after such quantum quenches can feature symmetry-broken phases and singular behaviour at the transition towards the 
disordered phase.  These Dynamical Phase Transitions (DPTs)~\cite{Sciolla,Sciolla_2,Silva} may be understood as transitions in the micro-canonical ensemble in case 
the many-body system thermalizes, driven by shifting the system's energy across the symmetry-restoration threshold. 
In non-ergodic systems, however, long-time steady states can be realized which cannot be described in terms of the conventional thermodynamic ensembles.
As a particular consequence, such systems allow the generation of phases and phase transitions with properties that cannot be realised in any equilibrium 
context~\cite{Huse2013}.  

In this work we focus on DPTs realized in spin-1/2 Ising models in transverse fields. We consider the case of long-range interacting models, which have recently
attracted a lot of attention~\cite{Piccitto_2019,Sciolla,Silva,PhysRevB.99.045128,PhysRevB.98.134303,PhysRevX.7.041021,PhysRevB.99.224203,
PhysRevLett.122.150601,guo2019signaling,PhysRevB.99.121112,verdel2019realtime} and constitute a paradigmatic class of non-ergodic systems capable 
of generating non-equilibrium steady states as reported both theoretically~\cite{Schiro2010,Sciolla,Sciolla_2,Silva} and experimentally~\cite{2017Neyenhuis,
Zhang2017,2017ZhangCrystal}. It was shown~\cite{Sciolla,Silva} that starting from an initial fully polarized state along the ordering direction, the asymptotic 
state of these systems can undergo a transition from an ordered phase at small fields to a disordered one when the field exceeds a critical value.
While inherently of non-equilibrium character, the resulting phases can be characterized by means of the conventional Landau paradigm via local order parameters.
Still, the understanding of the nature of the transition between the ordered and disordered phases has remained limited.
In particular, it is unclear to which extent these DPTs follow the general paradigm of continuous equilibrium transitions such as to whether they can be 
categorized in terms of universality classes and therefore whether the concepts of universality and scaling extend to this non-equilibrium dynamical regime.
{We remark that here we completely neglect the analysis of singular behaviours in the (infinite-size) time dynamics, which is another aspect of DPTs~\cite{Heyl_2018} 
with some connection with the symmetry-breaking behaviour~\cite{Silva}.}  

In this work we show that the DPTs after a quantum quenches in transverse-field Ising chains with power-law decaying interactions ($\sim r^{-\alpha}$) can 
feature scale invariance. We find evidence that the critical exponents of the DPT are universal over a large range of  interaction exponents $\alpha$.
Via finite-size scaling of the time-averaged longitudinal magnetization we identify the critical value of the field $h_c$ of the DPT and, in particular, determine 
the scaling exponents of the transition.
By studying the decay in time of the longitudinal magnetization we are able to put bounds to the values of $\alpha$ above which the ordered phase disappears.
{We can confirm the existence of two phases as long as $\alpha \lesssim 2$. The time-averaged magnetization decreases with the averaging time and never reaches a plateau. 
This indicates that only the trivial phase survives in this regime of $\alpha$ consistent with previous works~\cite{Silva}.} 

For $\alpha=0$ the dynamics can be solved via an effective mean field description, which becomes exact in the thermodynamic limit (see for example~\cite{Vidal_PRA04}).  For $\alpha\geq 0$ we compute 
the quantum real-time evolution by means of the Discrete Truncated Wigner Approximation (DTWA)~\cite{Wootters}. It has already been reported that DTWA compares 
well with other methods for long-range models~\cite{Schachenmayer,PhysRevB.98.134303} and, as we are going to show, works very well also for our problem, giving a very good 
comparison with the results of a recent numerical study using tensor network methods~\cite{Silva}. The DTWA has the advantage that it allows us to access large sizes 
with moderate computational resources polynomially scaling in the system size. Consequently, we can perform finite-size scaling also in long-range systems where it is 
crucial to reach large system sizes in order to tell the difference from the infinite-range ($\alpha =0$) case.

When analyzing scale invariance at the DPT, we find that the DTWA gives rise to scaling exponents identical to the mean-field ones at $\alpha=0$.
For finite $\alpha$, at the mean-field level, the exponents are of course independent on the range of the interaction. This is different for the DTWA, 
it compares well with exact methods, as emphasized above, and it is able to capture correlations. Therefore, in principle, it can give reliable scaling exponents.  
We computed the dependence on $\alpha$ of the scaling exponents of the magnetization,  and observed  a significant deviation from  the mean-field values at $\alpha \sim 1$. 
As discussed in the relevant sections, in this regime of $\alpha$ DTWA is not able to achieve accurate precision for a reliable scaling. It clearly indicates however when 
the deviations from mean field occur.

The favourable scaling of the DTWA with the number of sites allows to tackle the study {of} the DPT also in higher dimensions, a problem never touched so 
far in the literature. As long as the spins are interacting via long-range exchange couplings we do not expect significant dependence on the dimensionality. 
This is why we decided to study a two-dimensional system with nearest-neighbour coupling. Also in this case we expect a DPT. Here however, the critical 
behaviour should clearly deviate from the mean-field case. In this case we can only compare to exact diagonalization at small system sizes to test the quality of the DTWA 
approach. As we will show in the second part of the paper, we are able to detect the existence of the DPT through an analysis of the magnetization and 
of the Binder cumulant. We {perform a comparison with finite-temperature quantum Monte Carlo results and we }see that this DPT corresponds to the thermal transition. 
We show that this result is physically sound because the model is quantum chaotic. 
We find additional support for this conclusion {by using exact diagonalization and showing that the level-spacing statistics is Wigner-Dyson. } 

{In addition, we believe that our work may also contribute to a better understanding of the range and the limitations of
qualitative and quantitative applicability of the DTWA. The DTWA has been proved to work better in the context of long-range interactions~\cite{Schachenmayer}. The reason is that DTWA catches the long-distance quantum correlations only partially and then works better when the model is near to be infinite-range. The situation is similar to the one of the mean-field approximation, with the improvement that here quantum correlations are taken into account at least partially, giving rise to scaling exponents beyond the mean-field result. In the two-dimensional case quantum correlations become more relevant for the dynamics and we find that the DTWA provides only a qualitative (but remarkably meaningful) description for the dynamics.}

The paper is organized as follows. In Section~\ref{sec:model_methods} we introduce the model and also define the order parameter for the phase transition. 
In Section~\ref{DTWA:sec} we discuss the DTWA theory in detail and we show how to apply it to our model. In Section~\ref{comparison:sec} we compare 
the DTWA approach for this model with known results both in the infinite-range interaction case -- where exact diagonalization is possible also for large 
sizes -- and long-range interaction where the TDVP method is used. For the range of parameters we are interested in, we find that the comparison is very good. 
In Section~\ref{scaling:sec} we perform the finite-size scaling analysis for the one-dimensional long-range case. We first consider the case $\alpha=0$ where 
we compare with the exact diagonalization results and find that the comparison is very good. Then we move to analyze the case $\alpha\neq 0$ and see 
that the transition exists only  when $\alpha \lesssim 2$.  The results for the short-range two-dimensional models are discussed in Section~\ref{2dcase}.  Finally, 
Section~\ref{sec5} is devoted to the conclusions and further perspectives. The appendices contain additional details of the numerical analysis.

\section{The model }
\label{sec:model_methods}

As anticipated in the introduction, we will study a system of $N$ interacting spins governed by the Hamiltonian
\begin{equation}
\label{eq:Hamiltonian}
  	\hat{H}=- \sum_{i\neq j} J_{ij} \hat{\sigma}_{i}^{x}\hat{\sigma}_{j}^{x} -h\sum_{i}\hat{\sigma}_{i}^{z}~,
\end{equation}
where the $\hat{\sigma}_j^{x,z}$ are the Pauli matrices of the spin located in the $j-$th site, $h$ is an external transverse field and $J_{ij}$ is the 
exchange coupling between the spins. We will consider two cases (assuming to express the energies in units of the exchange coupling):

\begin{itemize}

\item 
A long-range interacting spin exchange 
\begin{equation}
\label{lr-J}
 	J_{ij}=  \frac{K_{\alpha}}{r_{ij}^\alpha} 
\end{equation}
in one dimension. We assume periodic boundary conditions and define the distance between two sites as $r_{ij}=\min[|i-j|,N-|i-j|]$. {The Kac factor~\cite{Kac} $K_\alpha$
 is defined as $K^{-1}_{\alpha} \equiv \frac{1}{N-1}\sum_{i\neq j}^{N} r_{ij}^{-\alpha}$ and} ensures that the Hamiltonian is extensive.

\item A short-range interacting spin on a $d$-dimensional cubic lattice where the exchange coupling
\begin{equation}
\label{sr-J}
 	J_{ij}=  \frac{1}{d}\delta_{i,\mbox{\small{nn}}(j)}
\end{equation}
is different from zero ($\delta_{l,m}$ is the Kronecker-delta) only if $i$ and $j$ are nearest-neighbours (nn). We will assume periodic boundary 
conditions and consider the cases of $d=1$ and $d=2$ (a square lattice of size $L$, $N= L^2$).
\end{itemize}

\vspace{0.5cm} 
 
The system is initialized  in the state fully polarized along  $x$, 
\begin{equation} 
	\label{psi0:eqn}
  	\ket{\psi_0}= \bigotimes_i \ket{\rightarrow}_i\,.
\end{equation}
We then perform a quantum quench {with the dynamics governed by the Hamiltonian} of Eq.~\eqref{eq:Hamiltonian}.
We are interested in the evolution of the total $x$ (longitudinal) magnetization which is given by
\begin{equation} \label{mixti:eqn}
  m_x(t)=\frac{1}{N}\sum_{i=1}^{N}\bra{\psi(t)}\hat{\sigma}_i^x\ket{\psi(t)}
\end{equation}
and the order parameter for the DPT is the long-time average of this magnetization. 
\begin{equation}
\label{eq3}
  \overline{m}_x=\lim\limits_{T \to \infty}\frac{1}{T}\int_{0}^{T}dt~m_x(t)~.
\end{equation}

({We will always use the finite-$T$ version of this quantity, $\overline{m}_x(T)$. We will not specify the dependence on $T$ in those cases where we have attained convergence.}). In the short-range two-dimensional case, we will also analyze the Binder cumulant in the long-time limit, defined as 
\begin{equation} \label{bionder:eqn}
	U_L = 1 - \overline{ \frac{ m^{(4)}_x }{ 3(m^{(2)}_x)^2} }
\end{equation} 
where we defined $m^{(l)}_x(t)=\frac{1}{N^l} \bra{\psi(t)} \left[\sum_{i=1}^{N}\hat{\sigma}_i^x\right]^l \ket{\psi(t)}$. 

The Binder cumulant is a measure for non-Gaussian
fluctuations of the order parameter. At equilibrium in the thermodynamic limit it acquires two different universal values in the two phases: The Gaussian value 0 in the disordered phase and the value $2/3$ in the ordered phase. At the transition point the Binder cumulant is scale invariant and it is a very convenient numerical probe for the existence of an equilibrium transition~\cite{binder}. We will show that also in this non-equilibrium context for the 2d short-range case it behaves in the same way and allows to probe the existence of a transition.

\section{Discrete truncated Wigner approximation} \label{DTWA:sec}
Before getting into the discussion of the results, it is useful to recap the basic ideas behind the DTWA and to discuss the accuracy of this 
method for this problem. In the following, we first review methodological details of the DTWA and afterwards we use exact diagonalization 
and {matrix product state descriptions by means of a time-dependent variation principle (MPS-TDVP)} data for a quantitative 
comparison.

\subsection{DTWA Method}

The DTWA is a semiclassical approximation, which has been used in many contexts concerning long-range interacting spin systems and has given 
noteworthy results, in terms of comparison with exact results and scalability to large system sizes. 
Precise details on the background can be found in~\cite{Silvia,Schachenmayer}, here we outline our concrete implementation. All the 
analysis is based on the construction of the discrete Wigner representation~\cite{Wootters} which is a 
generalization to discrete Hilbert space of the usual Wigner representation (details can be found in Ref.~\onlinecite{Polkovnikov}). 
Summarizing, Wootters has shown that, given a discrete Hilbert space, the quantum dynamics can be represented through a 
discrete basis of operators. In the case of a single $1/2$ spin, a possible basis choice is
\begin{equation} \label{ini:eqn}
  \hat{A}_{\beta}=\frac{\boldsymbol{1}+\textbf{s}_{\beta}\cdot\hat{\boldsymbol{\sigma}} }{2}
\end{equation}
where $\boldsymbol{s}_\beta$ can take the values $\left(\begin{array}{ccc}1&1&1\end{array}\right)$, $\left(\begin{array}{ccc}-1&1&-
1\end{array}\right)$, $\left(\begin{array}{ccc}1&-1&-1\end{array}\right)$ and $\left(\begin{array}{ccc}-1&-1&1\end{array}\right)$ and 
$\hat{\boldsymbol{\sigma}}=\left(\begin{array}{ccc}\hat{\sigma}^x&\hat{\sigma}^y&\hat{\sigma}^z\end{array}\right)$. 

With this basis choice, the expectation of any operator $\hat{\mathcal{O}}$ acting on the Hilbert space of the single spin can be written as
\begin{equation}
  \mean{\mathcal{O}}_t=\sum_{\beta}w_\beta\,\mathcal{O}_\beta(t)
\end{equation}
where $w_\beta\equiv\frac{1}{2}\Tr\left[\hat{A}_{\beta}\hat{\rho}\right]$ is the Wigner function, $\mathcal{O}_\beta^w(t)={\frac{1}{2}}\Tr\left[\hat{A}_{\beta}\hat{\mathcal{O}}(t)\right]$ 
are the Weyl symbols and $\hat{\mathcal{O}}(t)\equiv\nep^{i\hat{H}t}\hat{\mathcal{O}}\nep^{-i\hat{H}t}$. This representation can be extended also to our 
case of $N$ spins considering as basis operators
\begin{equation}
  \hat{A}_{\boldsymbol{\beta}}=\hat{A}_{\beta_1}\otimes\hat{A}_{\beta_2}\otimes\hat{A}_{\beta_3}\otimes\cdots\otimes\hat{A}_{\beta_N}
\end{equation}
and writing as before the expectation of any operator $\hat{\mathcal{O}}$ acting on the Hilbert space of the $N$ spins as
\begin{equation}
  \mean{\mathcal{O}}_t=\sum_{\boldsymbol{\beta}}w_{\boldsymbol{\beta}}\,\mathcal{O}_{\boldsymbol{\beta}}(t)\,.
\end{equation}
Up to now everything is exact. The DTWA amounts to approximate the time-evolved basis operators as factorized objects
\begin{equation}
  \hat{A}_{\boldsymbol{\beta}}(t)=\nep^{-i\hat{H}t}\hat{A}_{\boldsymbol{\beta}}\nep^{i\hat{H}t}\simeq \hat{A}_{\beta_1}(t)\otimes\hat{A}_{\beta_2}(t)\otimes\cdots\otimes\hat{A}_{\beta_N}(t)
\end{equation}
where
\begin{equation}
  \hat{A}_{\beta_j}(t)=\frac{\boldsymbol{1}+{s}_{j,\,\beta_j}^x(t)\hat{\sigma}_j^x+{s}_{j,\,\beta_j}^y(t)\hat{\sigma}_j^y+{s}_{j,\,\beta_j}^z(t)\hat{\sigma}_j^z }{2}\,.
\end{equation}
The ${s}_{\beta_j}^\nu(t)$ are initialized with the value for the corresponding $\beta_j$ given in Eq.~\eqref{ini:eqn} and obey a simple classical Hamiltonian dynamics given by
\begin{equation}
  \dot{s}_{j,\,\beta_j}^{\mu}(t)=\{s_{j,\,\beta_j}^{\mu}(t),{\cal H}\}=2\sum_{\nu\rho}\epsilon_{\mu\nu\rho}s_{j,\,\beta_j}^{\rho}(t)\frac{\partial{\cal H}}{\partial s_{j,\,\beta_j}^{\nu}}\,.
\end{equation}
Here the symbol $\{\cdots,\cdots\}$ is the Poisson bracket, $\epsilon_{\mu\nu\rho}$ is the Levi-Civita fully antisymmetric tensor, the variables $s_{j,\,\beta_j}^{\mu}$ obey 
the angular-momentum Poisson brackets $\{{s}_{j,\,\beta_j}^{\mu},\,{s}_{l,\,\beta_l}^{\nu}\}=\delta_{j\,l}\epsilon_{\mu\nu\rho}s_{j,\,\beta_j}^{\rho}$ and the classical 
effective Hamiltonian is defined as
\begin{equation}
  \mathcal{H}(\{s_{j,\,\beta_j}^{\mu}\})=-\sum_{i\neq j}^{N}J_{ij}{s_{i,\,\beta_i}^{x}s_{j,\,\beta_j}^{x}}-h\sum_{i}s_{i,\,\beta_i}^{z}~.
\end{equation}
For instance, the total longitudinal magnetization Eq.~\eqref{mixti:eqn} can be evaluated in the DTWA scheme as
\begin{equation} \label{mixtiDTWA:eqn}
  m_x(t)=\sum_{\boldsymbol{\beta}}w_{\boldsymbol{\beta}}\frac{1}{N}\sum_{i=1}^{N}s_{i,\,\beta_i}^x{(t)}\,.
\end{equation}
In this form it still unpractical from the numerical point of view because the index $\boldsymbol{\beta}$ runs over $4^N$ values, so the sum would be unfeasible for 
large system sizes. The solution comes from the relation $\sum_{\boldsymbol{\beta}}w_{\boldsymbol{\beta}}=1$, so, in the cases when 
$w_{\boldsymbol{\beta}}\geq 0$, it behaves as a probability distribution and it can be sampled through Monte Carlo sampling. With the initialization we choose 
we are in one of these lucky cases (see~\cite{Schachenmayer} for more details) and we can write Eq.~\eqref{mixtiDTWA:eqn} as the average over $n_r$ random 
initializations where each $s_{j,\,\beta_j}$ is initialized with probability $1/2$ in the condition $\left(\begin{array}{ccc}1&1&1\end{array}\right)$ and probability $1/2$ in 
the condition $\left(\begin{array}{ccc}1&-1&-1\end{array}\right)$. 

{We remark that this operation is a sample over an operator basis. Indeed, the initial density matrix can be written as $\hat{\rho}(0)=\bigotimes_j\hat{\rho}_j(0)$ with $\hat{\rho}_j(0)=\frac{1}{2}\left(\hat{A}_{\footnotesize\left(\begin{array}{ccc}1&1&1\end{array}\right)}+\hat{A}_{\footnotesize\left(\begin{array}{ccc}1&-1&-1\end{array}\right)}\right)$ and the two operators $\hat{A}_{\footnotesize\left(\begin{array}{ccc}1&1&1\end{array}\right)}$ and $\hat{A}_{\footnotesize\left(\begin{array}{ccc}1&-1&-1\end{array}\right)}$ are sampled with equal $1/2$ probability. Many possible choices of operator bases are possible, moving to each of these different representations by means of a unitary transformation. We provide an example of that in Appendix~\ref{samplingAPP}.}

Remarkably, the error bars do not scale with the system size, so this method is feasible also in the case of 
large systems. Moreover, results converge with a small number of randomness realizations ($n_r$); we show an example of this convergence 
in Appendix~\ref{samplingAPP}.  Unless otherwise specified here we use $n_r=504$. 

{Finally different sampling schemes, related to different choices of the operator in Eq.~\eqref{ini:eqn}, can be employed. In Appendix~\ref{samplingAPP} we 
briefly discuss these possible choices. All the results presented in the paper are essentially independent on the sampling method. Unless we specify otherwise, throughout the paper
we use the sampling scheme specified in Eq.~\eqref{ini:eqn}.}

In the following we are going to compare the DTWA method  with the results of other numerical methods in order to show its value also in our case.

\subsection{Comparison with other methods}  
\label{comparison:sec}

The comparison was done only in the case of one-dimensional power-law interaction. In this case, in addition to the possibility to have results from exact 
diagonalization (ED), it is possible to compare our data with tensor-network (the MPS-TDVP) results~\cite{Haegeman,Silva} for larger sizes.

\begin{figure}
  \centering
  \begin{tabular}{c}
    \begin{overpic}[width=71mm]{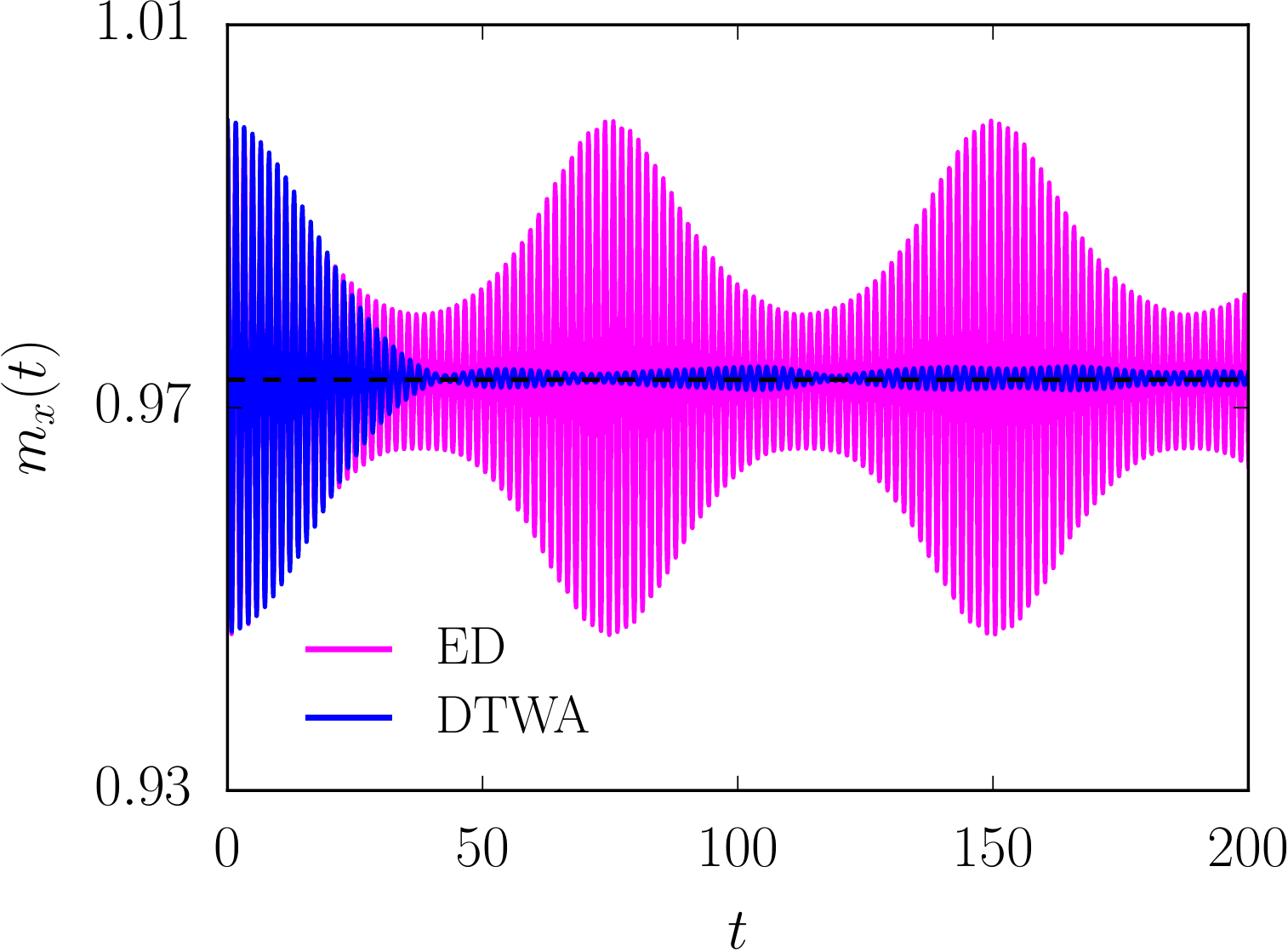}\put(-1,69){(a)}\put(30,67){$h=0.32$, $N=100$}\end{overpic}\\
    \begin{overpic}[width=71mm]{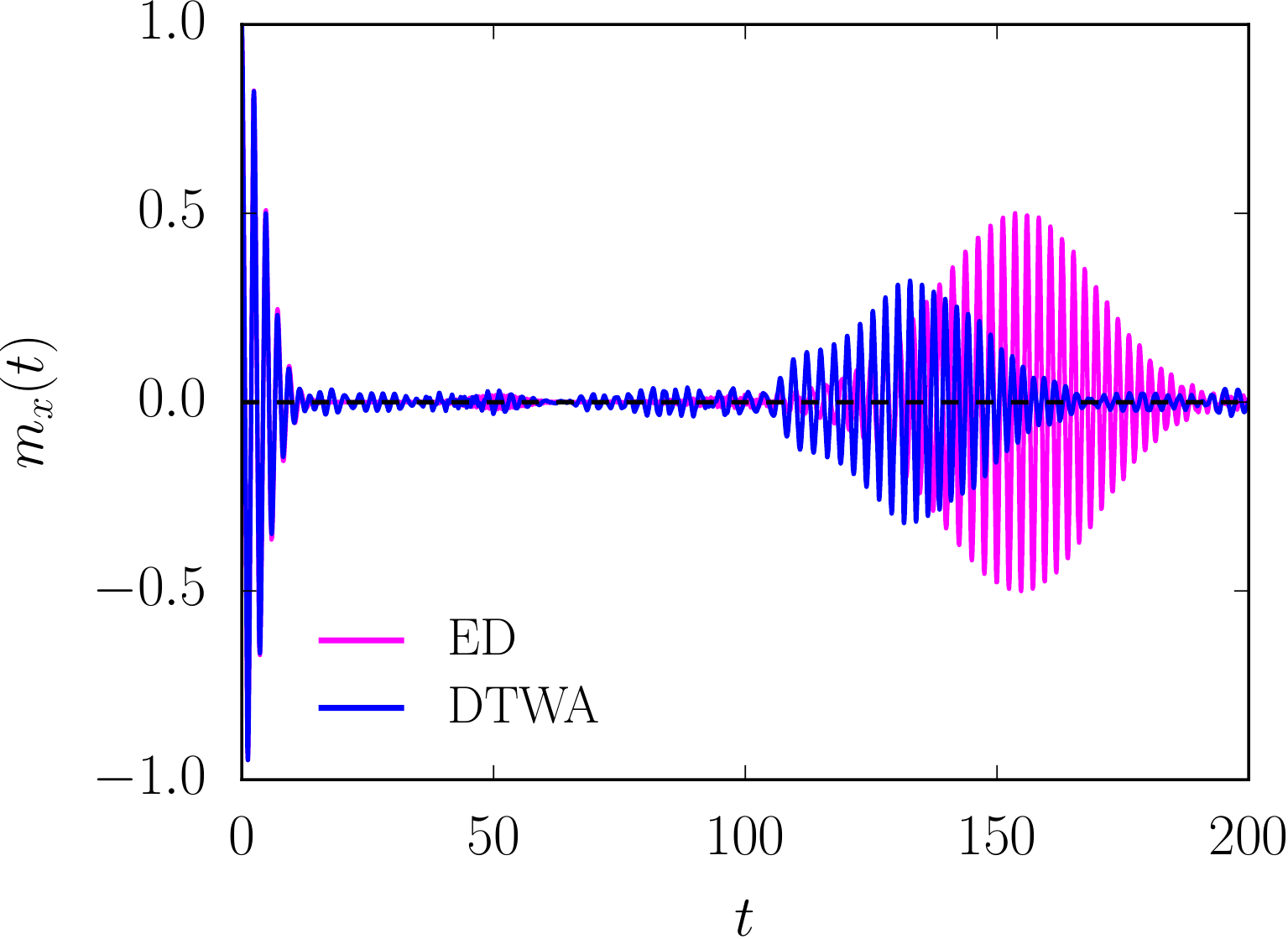}\put(-1,69){(b)}\put(26,61){$h=1.5$, $N=100$}\end{overpic}\\
    \hspace{0.5cm}\begin{overpic}[width=71mm]{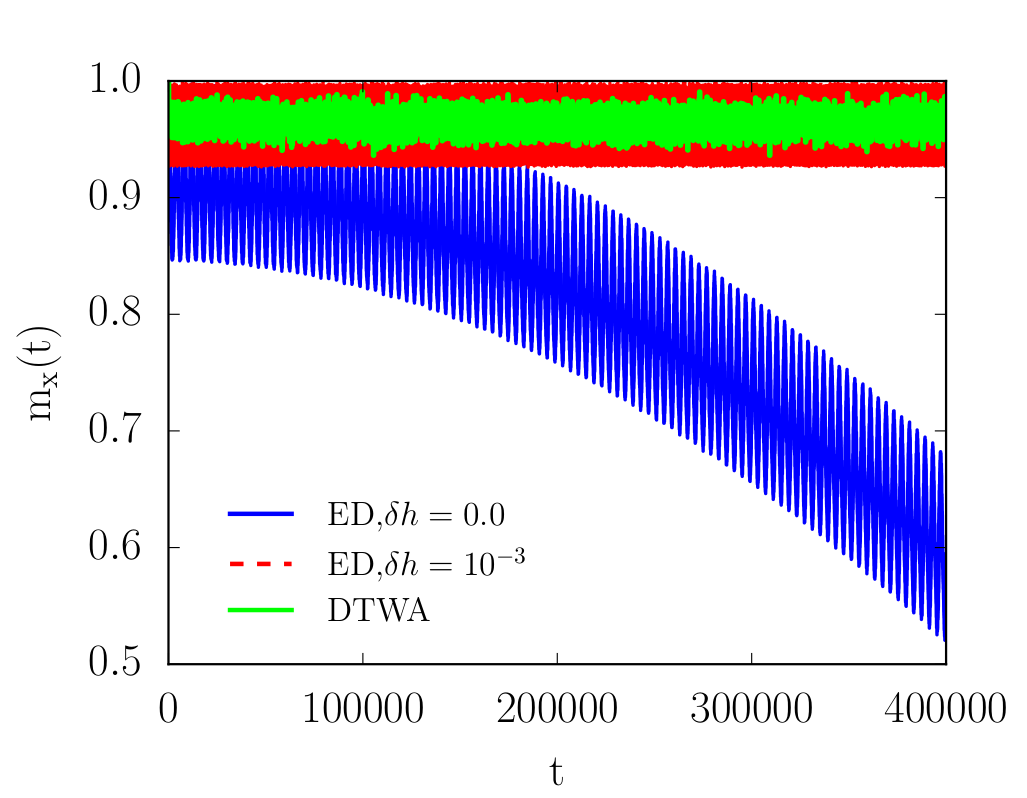}\put(-1,69){(c)}\put(20,35){$h=0.32$, $N=10$}\end{overpic}
  \end{tabular}
  \caption{Comparison of DTWA method with ED for different parameters. 
In panels (a) and (b) we can see that the ED and DTWA
curves loose agreement after a while but their time averages coincide
(dashed horizontal line).
DTWA cannot catch the Rabi oscillations; they are eliminated by adding a small symmetry-breaking field in the ED case 
  (see panel (c)). In panel (a) we use the sampling scheme specified by Eq.~\eqref{ini:eqn1} while in the other panels we use the sampling scheme given by Eq.~\eqref{ini:eqn}.}
	\label{fig:compare_ED}
\end{figure}
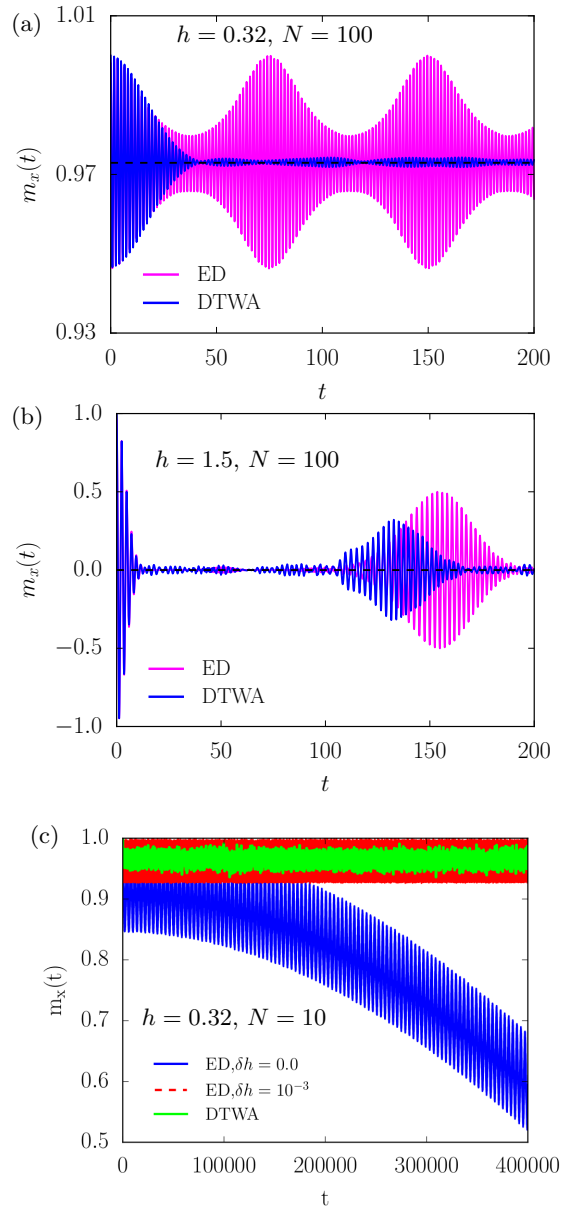

{First of all we consider the case $\alpha=0$ of infinite-range interactions. In this limit the model reduces to the Lipkin-Meshkov-Glick model whose exact diagonalization dynamics can be easily studied. With all the site-exchange operators conserved, there is a superextensive number of constants of motion and the dynamics becomes integrable. Thanks to the conservation of the modulus of the total spin, the quantum dynamics is restricted to a Hilbert subspace whose dimension scales linearly with the system size making the solution of large system sizes feasible. Specifically, the Hamiltonian commutes with the total-spin operator $\hat{\bf S}^2$ ($\hat{\bf S}=\frac{1}{2}\sum_j\hat{\boldsymbol{\sigma}}_j$ with $\hat{\boldsymbol{\sigma}}_j\equiv\left(\begin{array}{ccc}\hat{\sigma}_j^x&\hat{\sigma}_j^y&\hat{\sigma}_j^z\end{array}\right)^T$) and we can restrict to the $\hat{\bf S}^2$-subspace with eigenvalue $S(S+1)$ with $S=N/2$, which has a dimension $N+1$. (For a detailed explanation see for instance~\cite{angelo_th}). 
We show some instances of comparison in Fig.~\ref{fig:compare_ED}. Let's first consider the case $N=100$. We see that the curves of $m_x(t)$ deviate 
quite soon from each other, both for $h<1$ and $h>1$, but the time average (the one we are interested in) is actually the same (it is marked in the plots by a dashed horizontal line). }
{Dynamics up to a time $t\sim30$ is quantitatively correct. For larger times the quantum revivals are not captured properly. This feature, however, shifts to larger times upon increasing system size. Thus, for large systems this discrepancy becomes less and less relevant, making a description via the DTWA more accurate.}

We show also results for $N=10$. Here we can see that in the ED case a phenomenon appears which is not captured
by DTWA, the Rabi oscillations. Indeed, in this system an extensive number of eigenstates breaks the $\mathbb{Z}_2$ symmetry in the thermodynamic limit. 
For any finite size, the true eigenstates are the even and odd superposition of these symmetry-breaking states and are separated by an exponentially small 
gap. Preparing the system in a symmetry-breaking state (as the one in Eq.~\eqref{psi0:eqn}) gives rise therefore to Rabi oscillations of the magnetization 
with a frequency equal to the gap. {Because this gap is }exponentially small in the system size, we cannot see these oscillations in Fig.~\ref{fig:compare_ED}(a), where the size is 
$N=100$ and the gap is {negligibly} small {($\sim\nep^{-100\log(1/0.32)}$)}. But we can see them in Fig.~\ref{fig:compare_ED}(c) and they are not caught by DTWA. 

The existence of the Rabi oscillations is intimately related to the existence 
of a $\mathbb{Z}_2$ symmetry, and the presence of resonant symmetry-breaking states put in interaction by the term with the $h$ field.  Explicitly breaking 
the symmetry breaks the resonance and there are no more oscillations. We do this in Fig.~\ref{fig:compare_ED}(c) where we show also a curve of $m_x(t)$ obtained adding 
to the Hamiltonian a small symmetry breaking term $\delta h\sum_j\hat{\sigma}_j^x$. We see that there are no Rabi oscillations and the comparison 
with DTWA in terms of average is very good. So, in some sense, in DTWA one implicitly adds to the Hamiltonian a small symmetry breaking term. This 
is just what we operatively do when we want to see a quantum phase transition. We add a small symmetry-breaking term, we go to the thermodynamic 
limit, and then we send the small symmetry-breaking term to 0. Because we are interested here in the existence of a dynamical quantum phase transition 
with $\mathbb{Z}_2$ symmetry breaking, this is exactly what we should do. DTWA does this implicitly for us, and in the thermodynamic limit the presence 
of a small symmetry-breaking term makes no difference both for DTWA and ED.

 \begin{figure}
	\centering
	\begin{tabular}{c}
		\begin{overpic}[width=71mm]{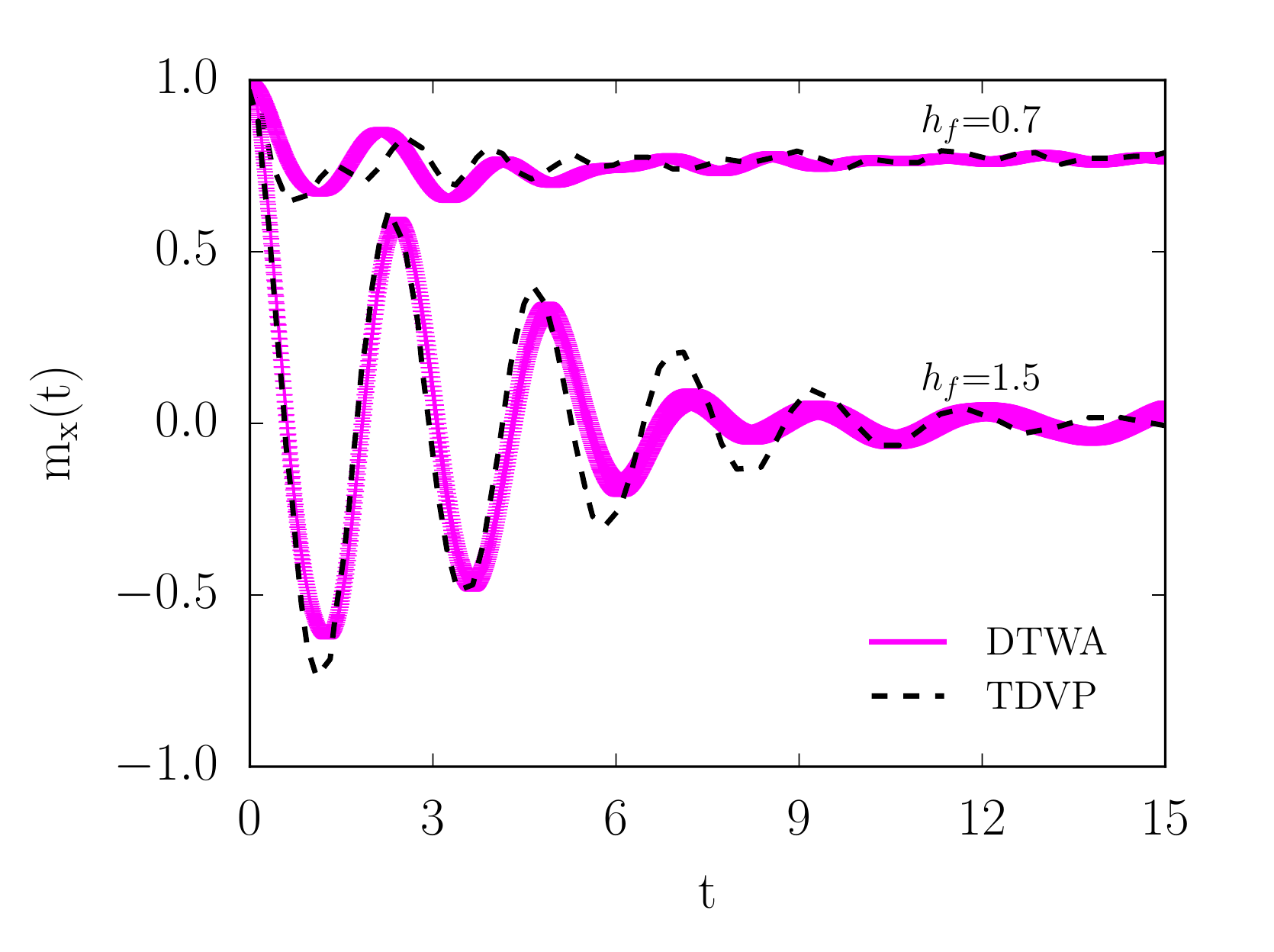}\put(-1,69){(a)}\put(30,20){$\alpha=1.5$}\end{overpic}\\
		\begin{overpic}[width=71mm]{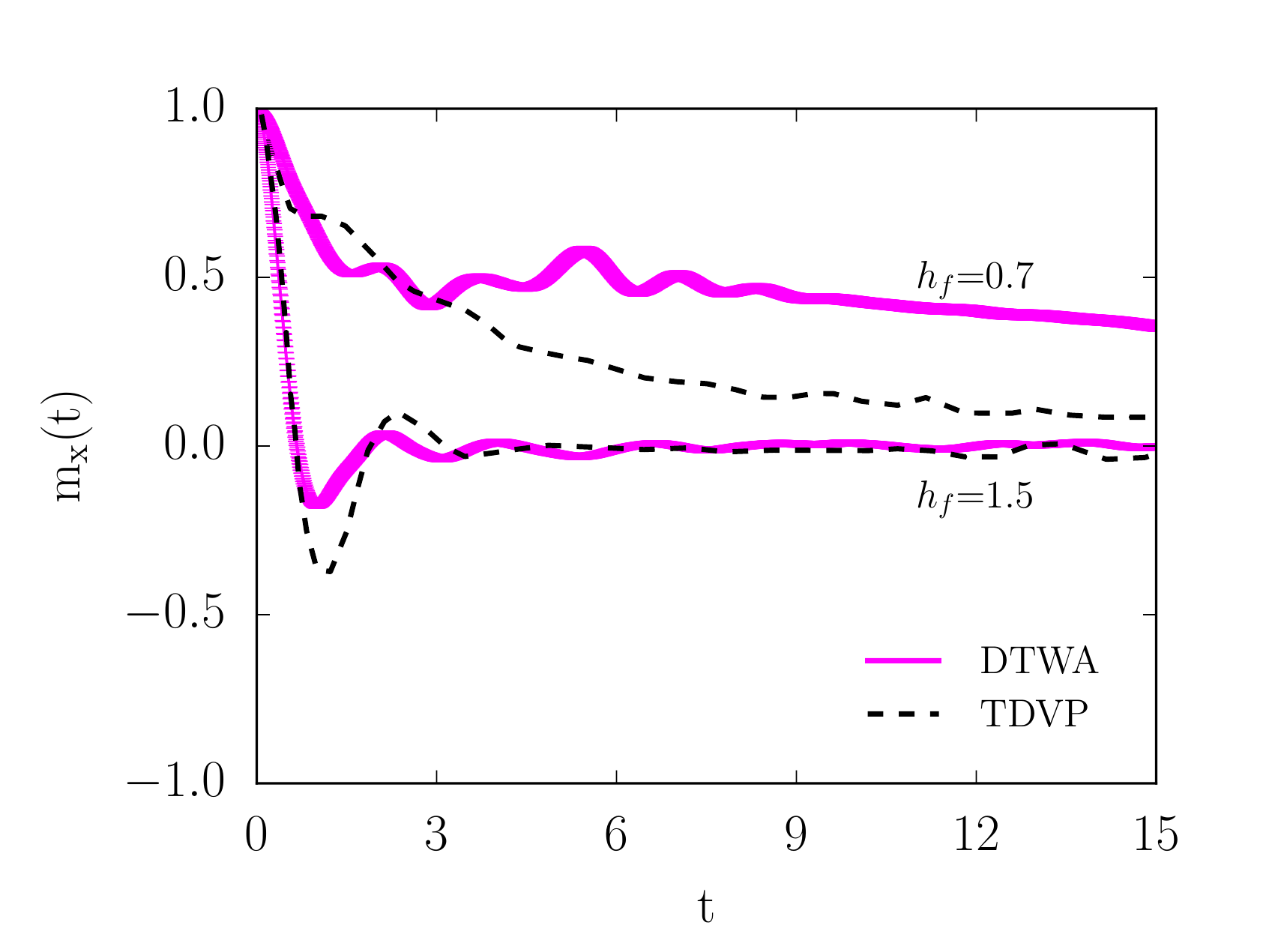}\put(-1,69){(b)}\put(30,20){$\alpha=3.0$}\end{overpic}
	\end{tabular}
	\caption{{The instantaneous magnetization} $m_x(t)$ versus $t$: Comparison of the results obtained with DTWA and TDVP~\cite{Silva} methods for different values of $\alpha$ and $h$. As expected from~\cite{Schachenmayer}
	 we see a much better agreement at smaller $\alpha$. Other parameters: $N=100$.}
	\label{fig:compare_TDVP}
\end{figure} 

\begin{figure}
	\centering
	\begin{tabular}{cc}
		\begin{overpic}[width=75mm]{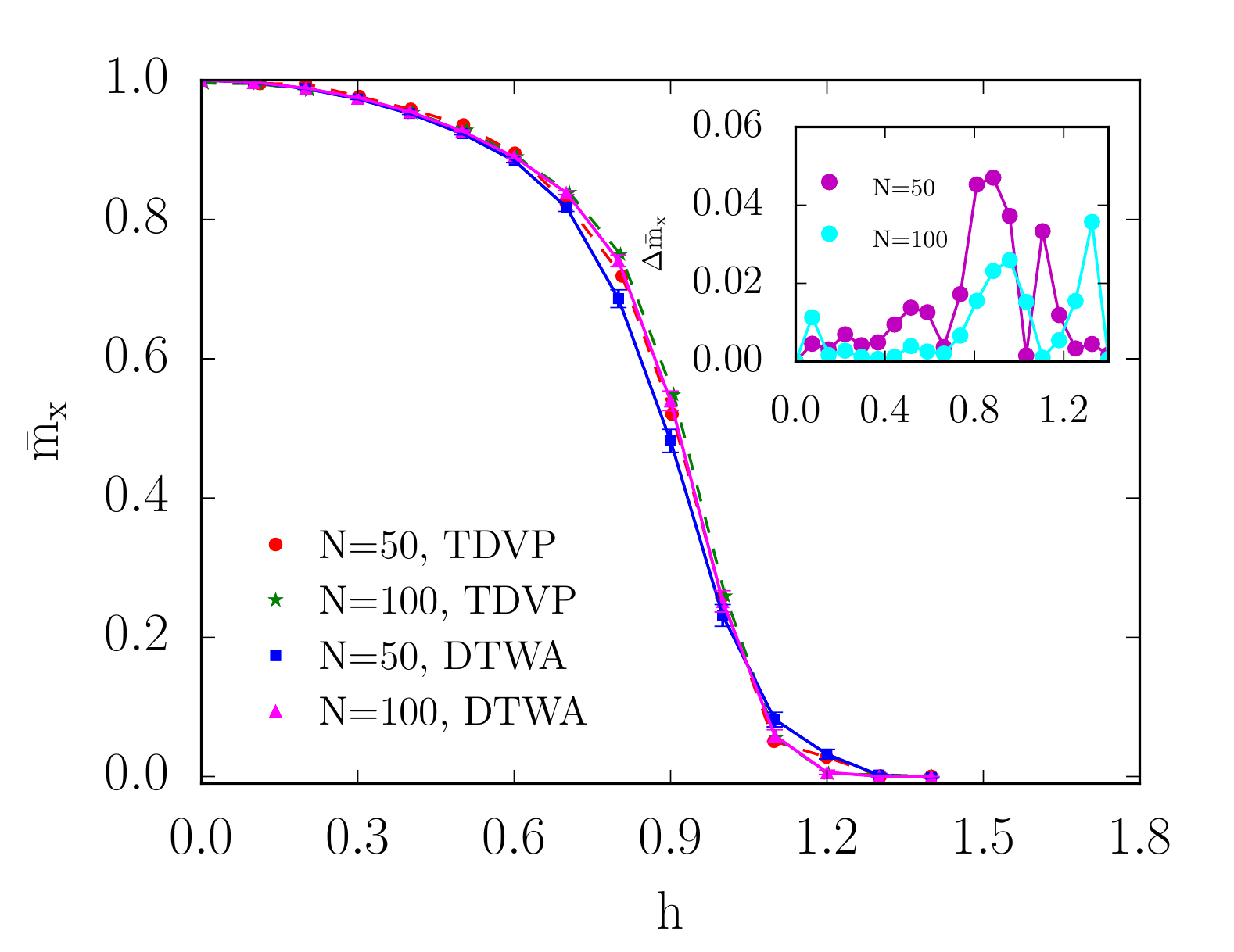}\put(-1, 69){(a)}\end{overpic}\\
		\begin{overpic}[width=75mm]{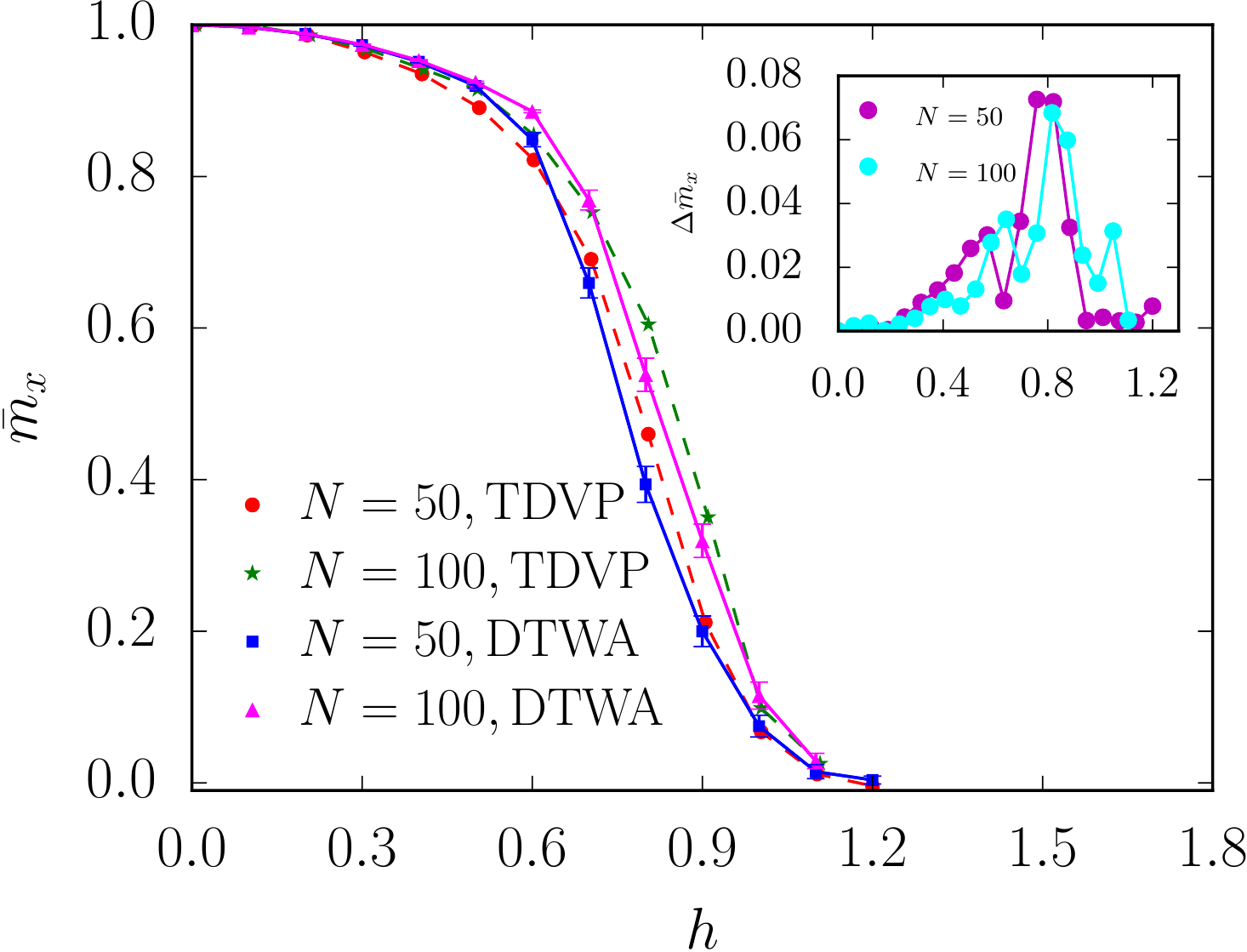}\put(-1,69){(b)}\end{overpic}
	\end{tabular}
    \caption{{The long-time average of the magnetization} $\overline{m}_x$ versus $h$: Comparison of the results obtained with TDVP~\cite{Silva} and DTWA methods for (a) $\alpha=0.1$ and (b) $\alpha=1.5$. 
    The insets indicate the difference between two methods where $\Delta \overline{m}_x=\abs{\overline{m}_x^{DTWA}-\overline{m}_x^{TDVP}}$.}
	\label{fig:DPD_silva}
\end{figure}

For $\alpha\neq 0$ we can compare our DTWA results for the transverse magnetization $m_x(t)$ with the corresponding ones obtained through the 
TDVP method~\cite{Haegeman,Silva} (see Fig.\ref{fig:compare_TDVP}) for the case of $N=100$ sites. The time scales we consider are 
much shorter than the times exponential in $N$ needed for seeing the Rabi oscillations. Let us start focusing on the case $\alpha=1.5$ 
[Fig.~\ref{fig:compare_TDVP}(a)]. We see that in this case the agreement is quite good both inside the symmetry-breaking phase ($h=0.7$) and outside 
it ($h=1.5$). On the opposite, for $\alpha=3$ [Fig.~\ref{fig:compare_TDVP}(b)] the agreement is very good only when $h=1.5$. When $h=0.7$, the 
DTWA result decays much more slowly than TDVP.  
The two methods are in agreement for small values of $\alpha$, as we expected from the existing literature on DTWA. In order to show the very good agreement 
when $\alpha$ is small, we plot in Fig.~\ref{fig:DPD_silva} the time-averaged longitudinal magnetization $\overline{m}_x$ versus $h$ for $\alpha=0.1$ 
and $\alpha=1.5$ obtained through the two methods. In both cases we see a very good agreement between the two methods. So, in the small-$\alpha$ 
regime we are interested in, the DTWA compares very well with the known 
results obtained through TDVP. This gives us an opportunity, because while TDVP can be used for at most $N=200$ (see~\cite{Silva}), 
DTWA can be pushed up to much larger sizes, thus offering the possibility of an accurate finite-size scaling.

\begin{figure}
	\centering
	\begin{tabular}{cc}
		\includegraphics[width=7cm]{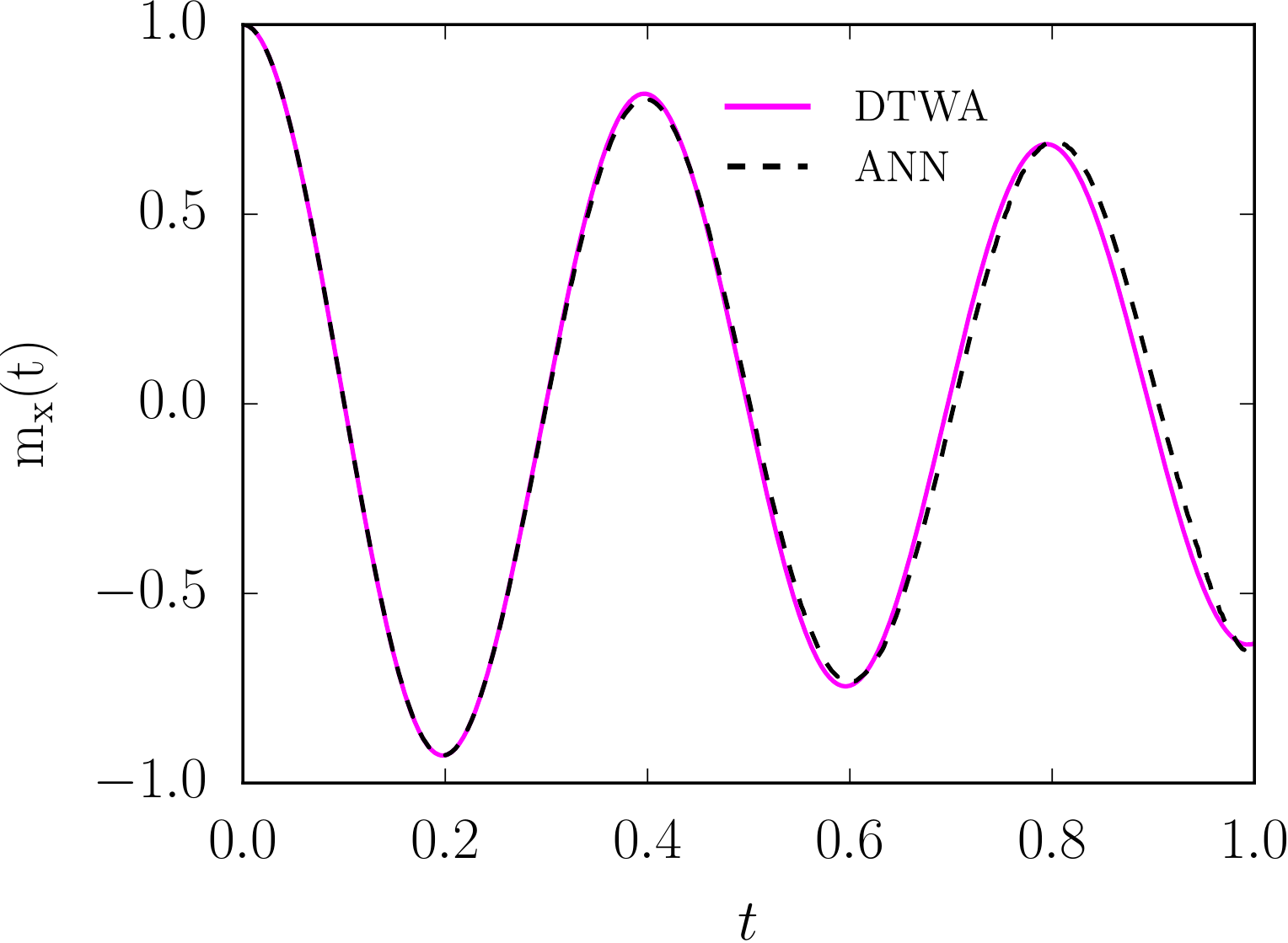}
	\end{tabular}
    \caption{Comparison of ${m}_x(t)$ obtained by DTWA with the same quantity obtained with ANN~\cite{schmitt2019quantum} for a two-dimensional short-range case. Numerical parameters: $h=8$ and  $n_r=10000$.}
	\label{fig:ANN}
\end{figure}

We conclude this section by comparing DTWA results for the two-dimensional short-range case with the dynamics obtained by means of artificial neural networks (ANN)~\cite{schmitt2019quantum}.
We show an example of comparison in Fig.~\ref{fig:ANN} with data taken from Ref.~\cite{schmitt2019quantum} at large transverse fields in the regime where the Ising symmetry is restored in the long-time limit. As one can see, the DTWA compares {remarkably }well with the numerically exact ANN data. {The idea of ANN approach is to encode the quantum many-body wave function in an artificial neural network~\cite{CarleoTroyer}. Importantly, ANNs are universal function approximators, which guarantees that the encoding always becomes asymptotically exact in the limit of sufficiently large ANNs. For the curve in Fig.~\ref{fig:ANN} it has been shown that the data has been converged with the size of the neural network, the result is indeed numerically exact.}

\section{Results} 
In this Section we will illustrate our results for the DPT obtained through the DTWA. We first analyze the one-dimensional  
long-range case [see Eq.~\eqref{lr-J}]. Later we will analyze the two-dimensional case with short-range interaction, Eq.~\eqref{sr-J}.
In this second case, {we  use also the Binder cumulant to get more reliable indications of the DPT}. In both cases we 
address the steady state properties, and consider the behaviour of the time-averaged magnetization~\eqref{eq3}. We consider averages over a time 
$T$ such that the magnetization has already converged and we specify it in any of the considered cases, explicitly studying the convergence in $T$ 
for $\alpha\gtrsim 2$.
\begin{figure}
	\begin{tabular}{c}
		\begin{overpic}[width=61mm]{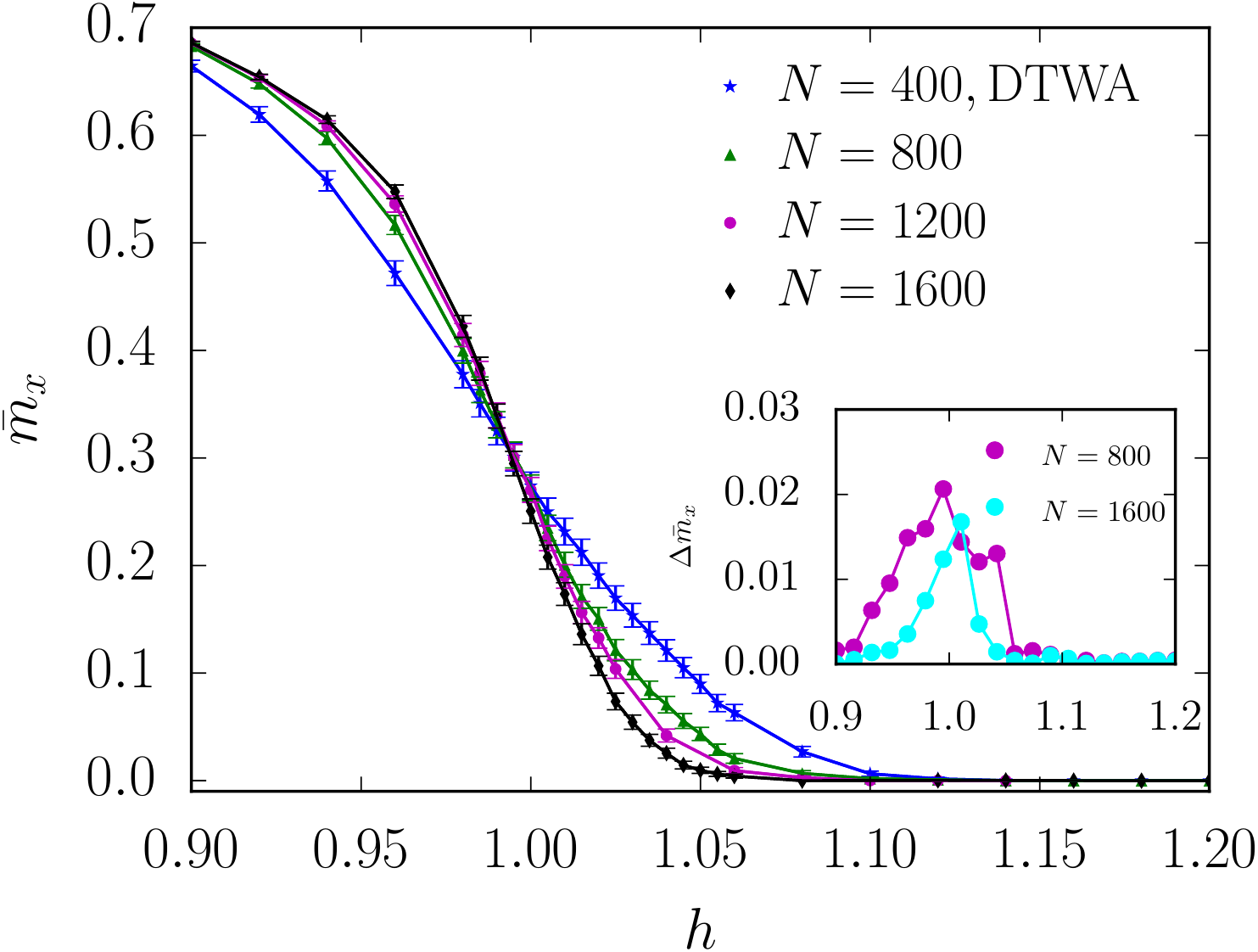}\put(22,25){(a)}\end{overpic}\\
                \begin{overpic}[width=61mm]{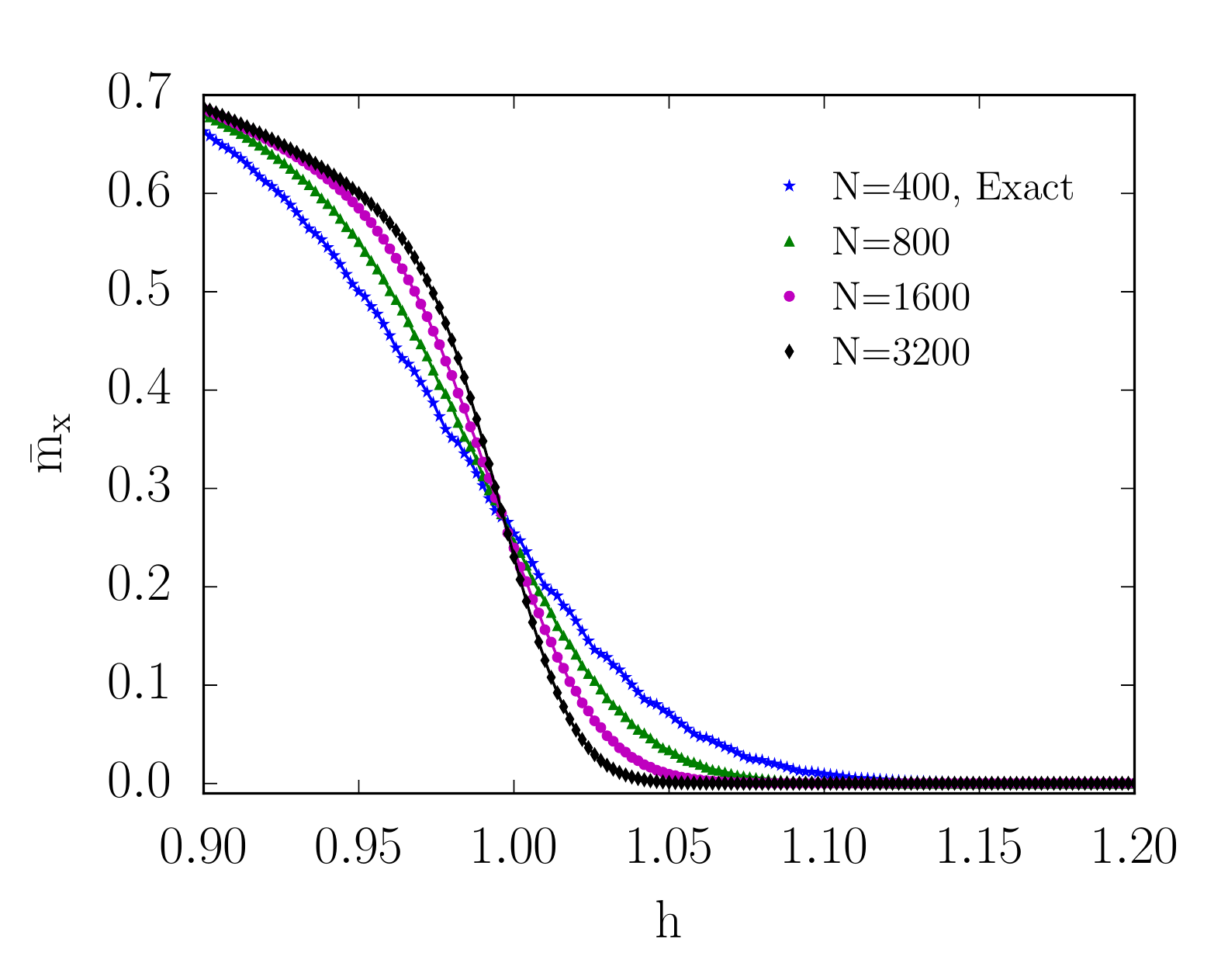}\put(22,25){(b)}\end{overpic}\\
	\end{tabular}
    \caption{The long-time average of the magnetization $\overline{m}_x$ versus the transverse field $h$ for $\alpha=0$ computed using the  
    DTWA [panel (a)] and exact diagonalization [panel (b)]. There is a {good} quantitative agreement  of the two data sets {(see the inset showing $\Delta \overline{m}_x=\abs{\overline{m}_x^{DTWA}-\overline{m}_x^{\rm Exact}}$ versus $h$ for two values of $N$)}. In both panels we perform the time average over $T=200$.}
	\label{plotmm:fig}
\end{figure}

\subsection{Long-range model in one dimension} 
\label{scaling:sec}

Let us first analyze the one-dimensional long-range case and study the finite-size scaling of $\overline{m}_x$ as a function of the transverse field $h$. We start 
with the case $\alpha=0$, where we can compare DTWA with the exact diagonalization. In Fig.~\ref{plotmm:fig} we plot the curves of $\overline{m}_x$ 
versus $h$ for different system sizes $N$, obtained through {DTWA [panel (a)] and exact solution [panel (b)]~\cite{note_th}.} {There is a good agreement}, comparing quantitatively very well, {as we show in the inset where we plot $\Delta \overline{m}_x=\abs{\overline{m}_x^{DTWA}-\overline{m}_x^{\rm Exact}}$ versus $h$ for two different values of $N$}.
As the system size is increased  both have one common crossing point $h_c$, making the existence of a phase transition clearly visible. Close to the crossing point 
the curves obey a scaling form of the type
\begin{equation} 
	\label{scalingb:eqn}
  	\overline{m}_{x,\,N}(h)=N^{-\beta}f\big[(h-h_c)N^{\delta}\big]~.
\end{equation}
The possible value $\beta \sim 0$ implies logarithmic corrections of the form $ \overline{m}_{x,\,N}(h) \sim (1/ \log  N )f( \cdot)$.

\begin{figure}
	\centering
	\begin{tabular}{c}
        \begin{overpic}[width=65mm]{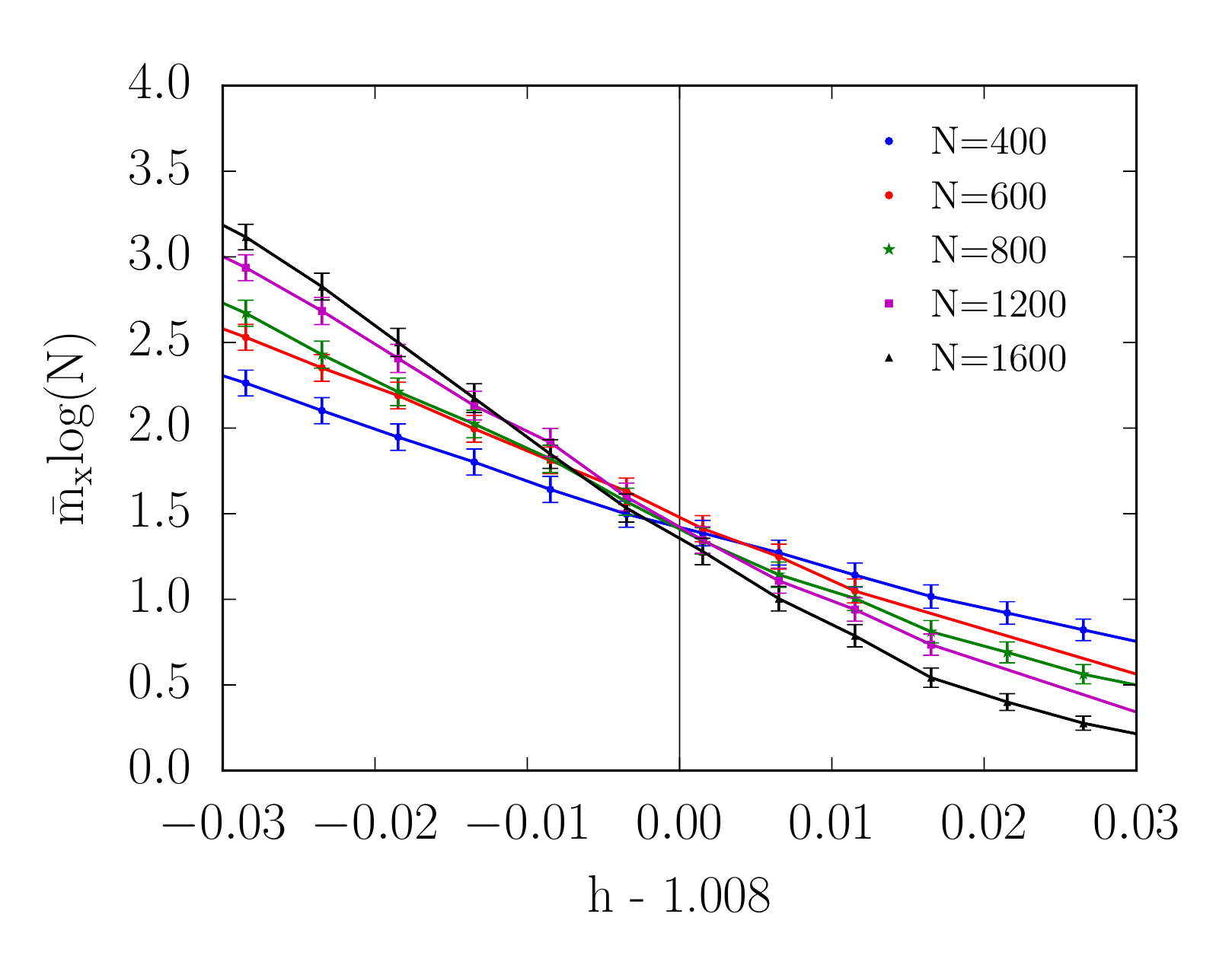}\put(22,25){(a)}\end{overpic}\\
            \begin{overpic}[width= 65mm]{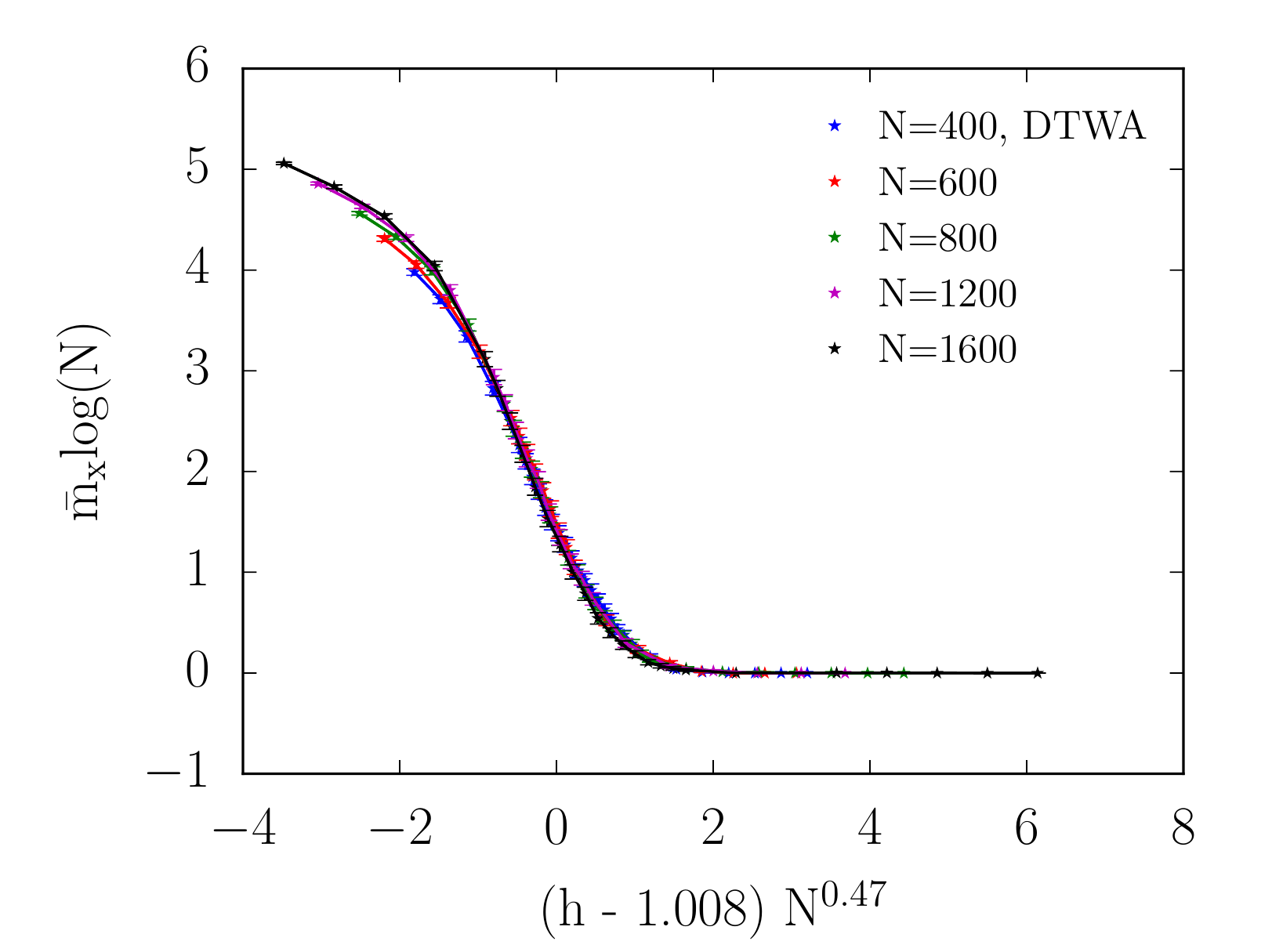}\put(22,25){(b)}\end{overpic}\\
                \begin{overpic}[width=65mm]{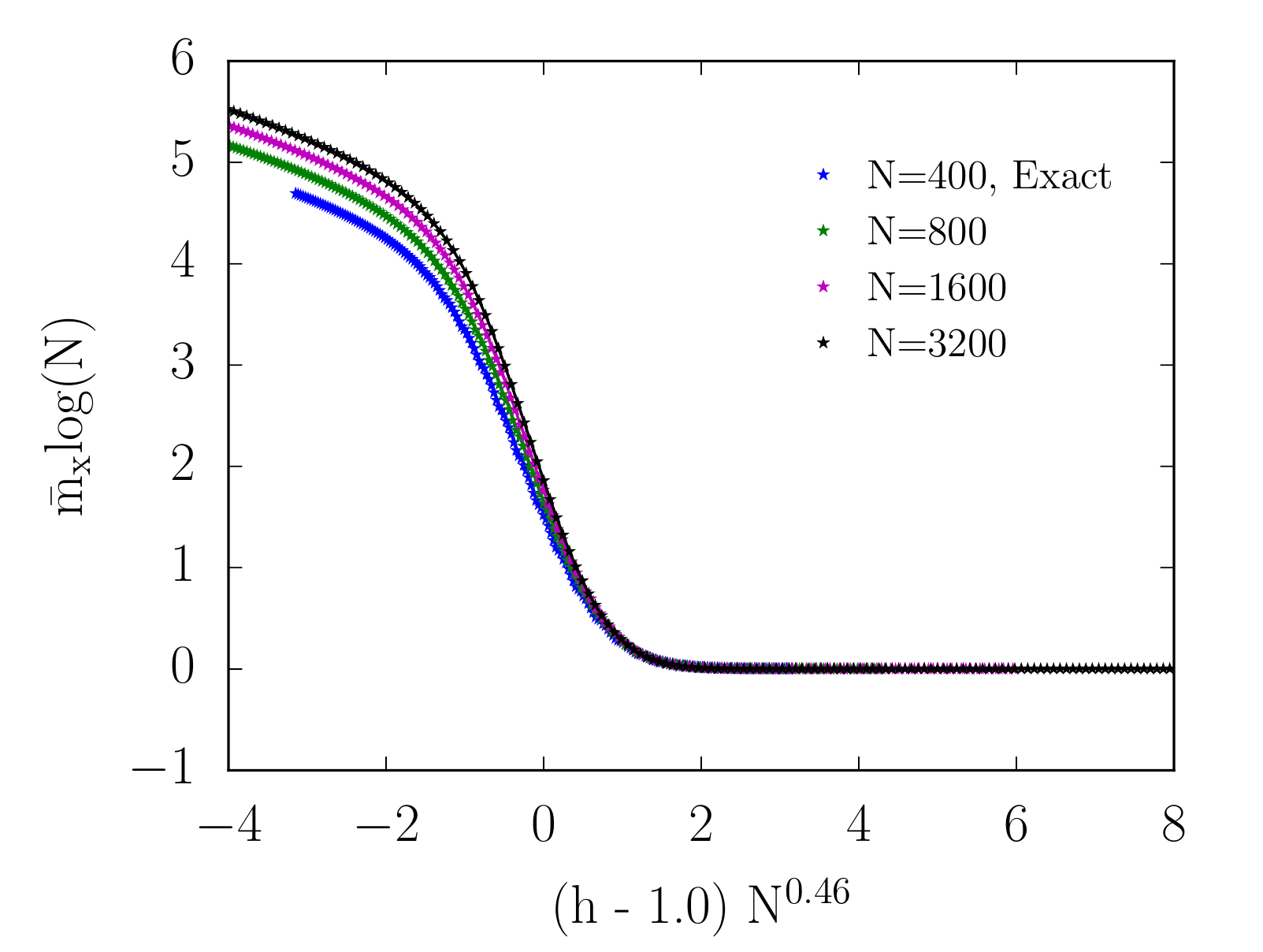}\put(22,25){(c)}\end{overpic}
	\end{tabular}
	\caption{($\alpha=0$) In panel (a) the data of  Fig.~\ref{plotmm:fig}(b) are magnified around the crossing region in order to see the details of the crossing region. 
	The imperfect crossing sets an error on the determination of critical field of the order of $10^{-2}$. Panels (b) and (c): Figs.~\ref{plotmm:fig}(a) and (b) rescaled 
	according to Eq.~\eqref{scalingb:eqn} with the choice of the optimal parameters. The scaling exponents essentially coincide in the two cases.}
	\label{plotmm_res:fig}
\end{figure}

\begin{figure*}
\centering
  \begin{tabular}{ccc}
      \begin{overpic}[width=55mm]{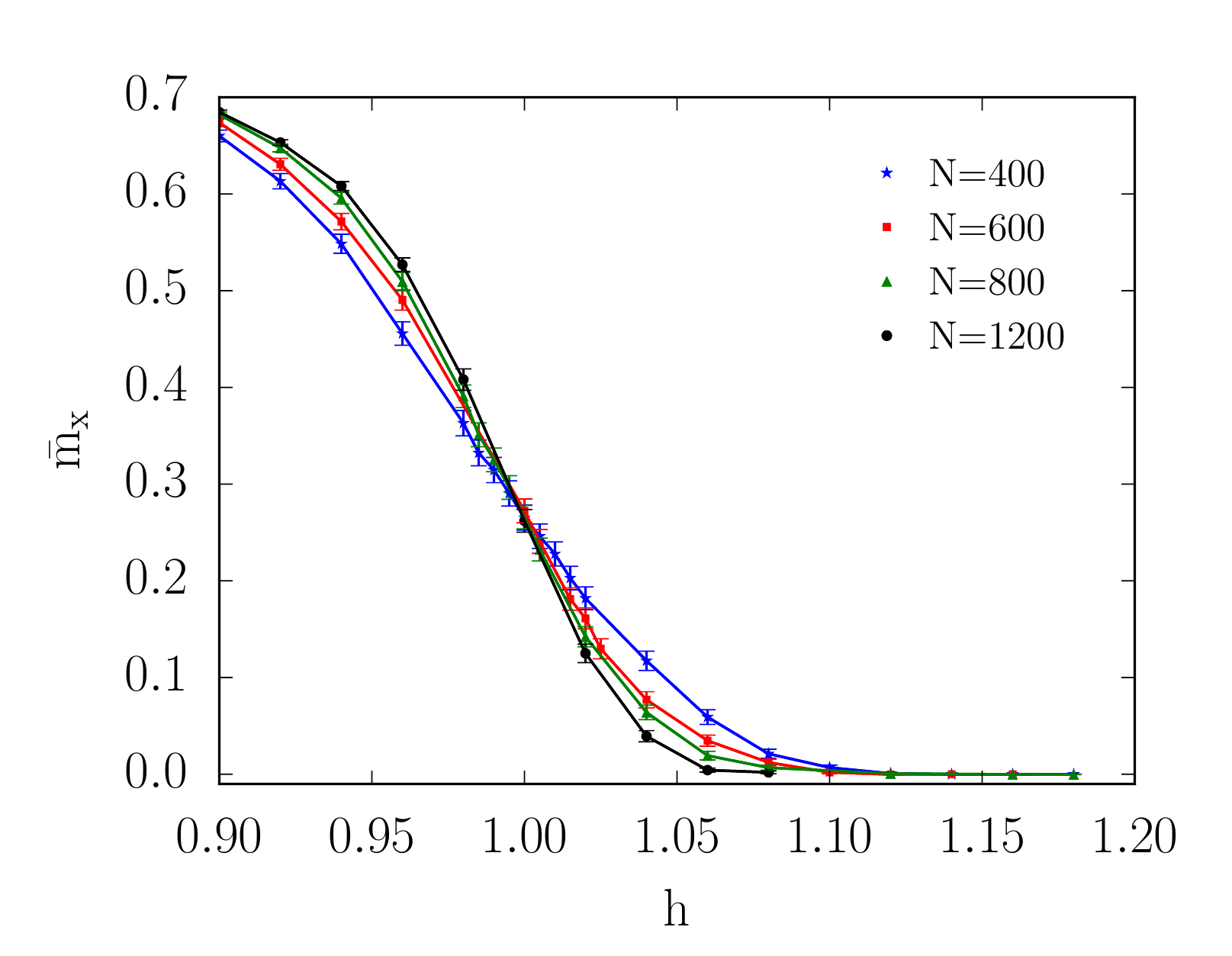}\put(-1,69){(a)}\put(40,61){$\alpha=0.1$}\end{overpic}&
          \begin{overpic}[width=55mm]{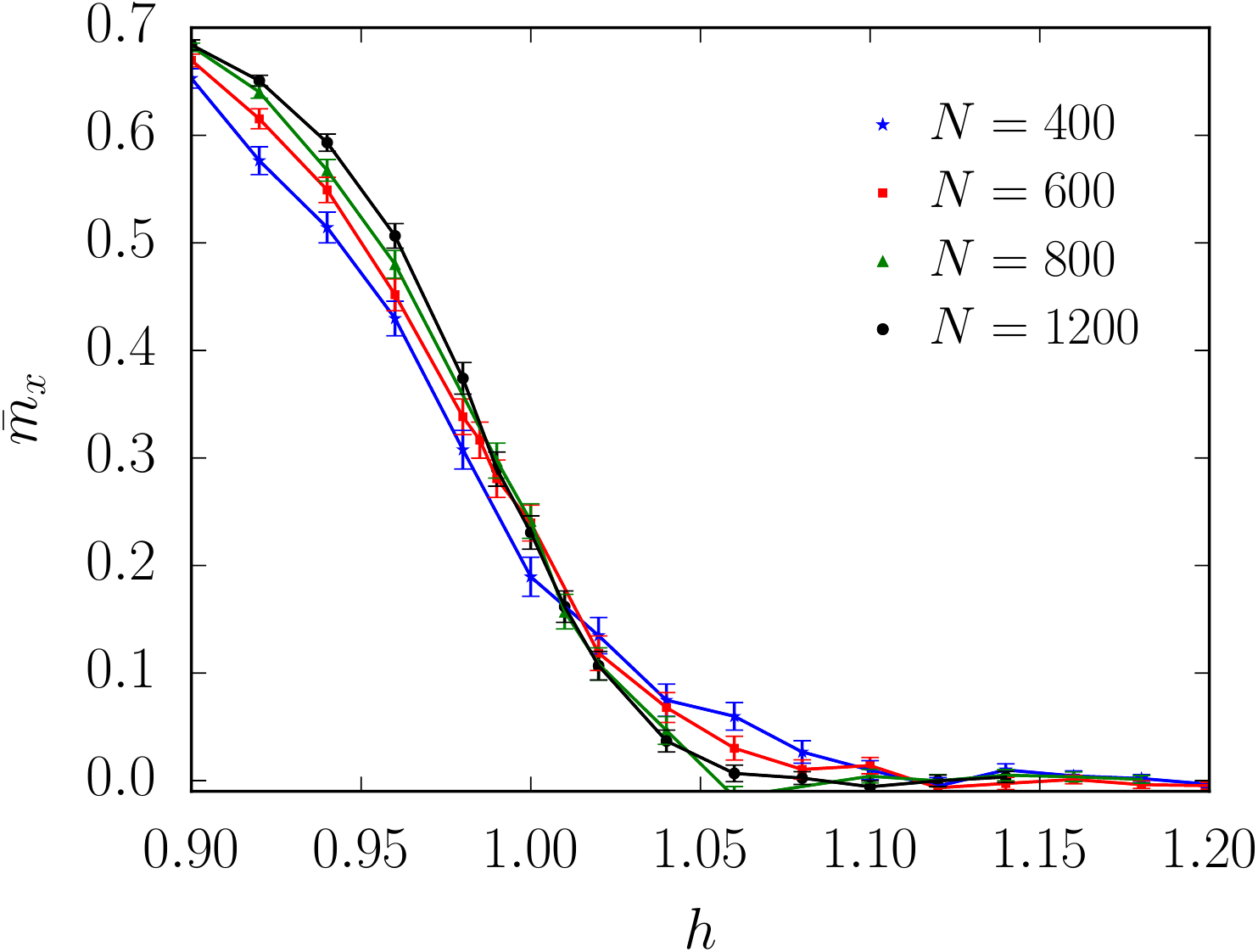}\put(40,61){$\alpha=0.5$}\put(-1,69){(b)}
        \end{overpic}\\
      \begin{overpic}[width=55mm]{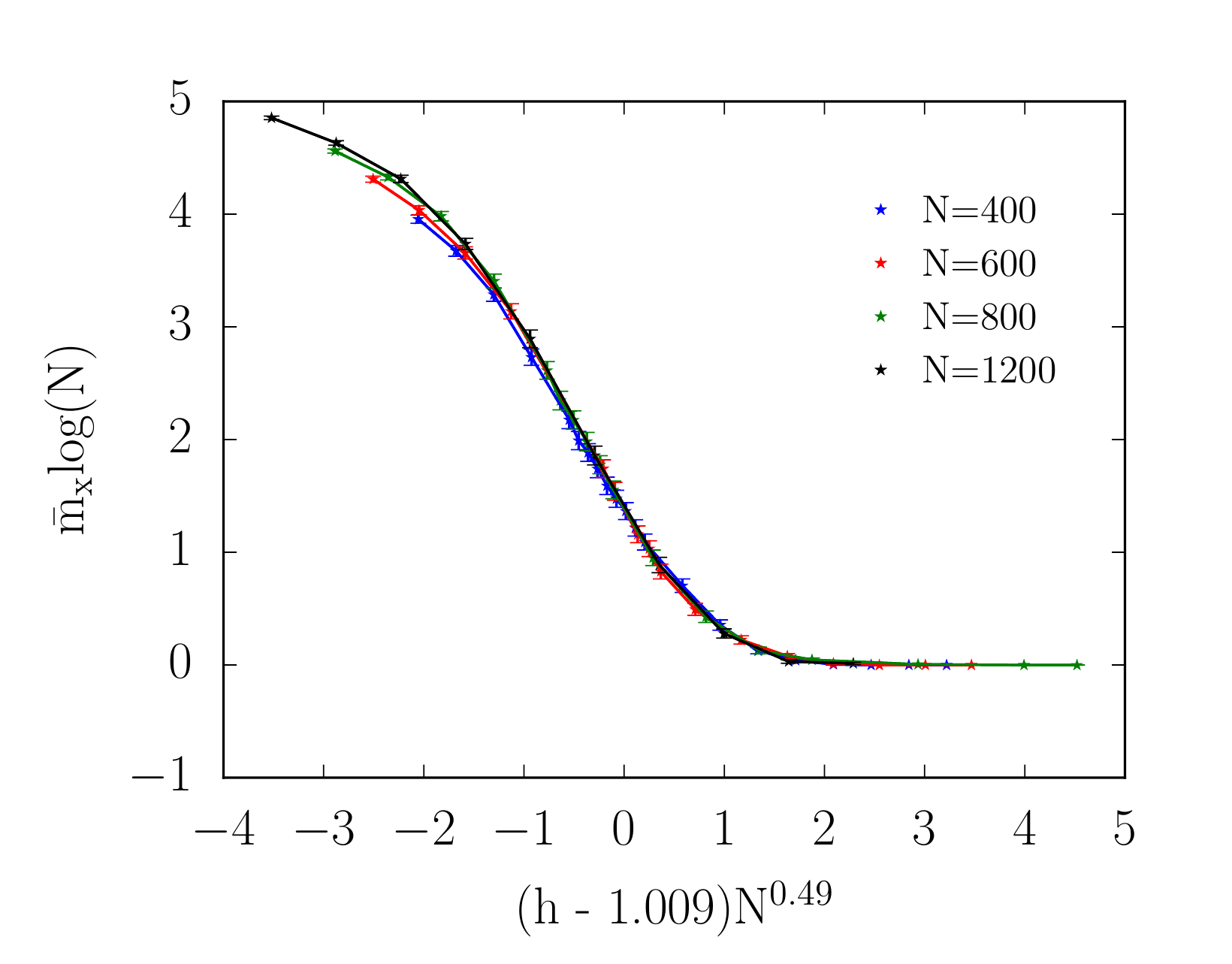}\put(-1,69){(c)}\put(45,65){$\alpha=0.1$}\end{overpic}&
          \begin{overpic}[width=55mm]{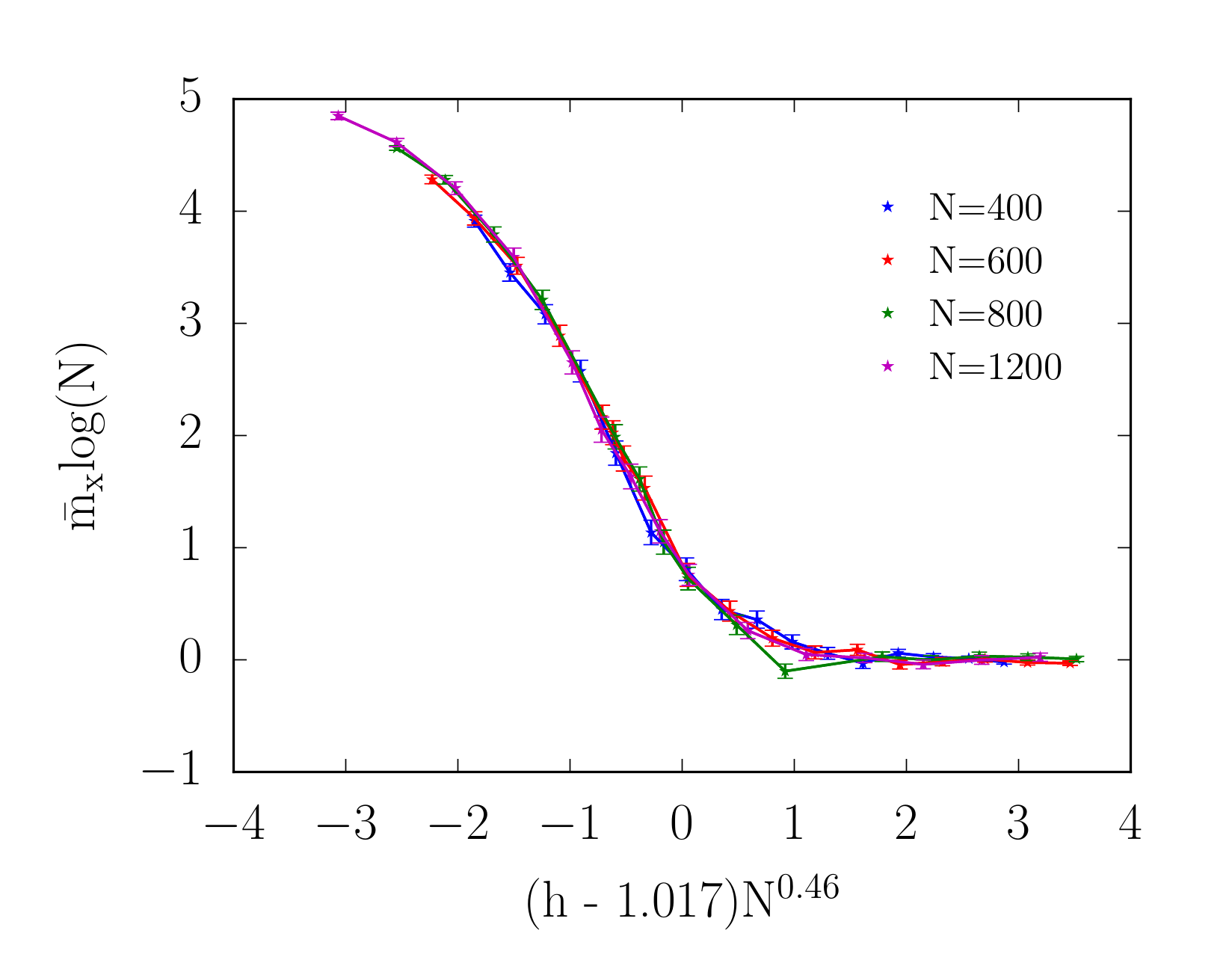}\put(-1,69){(d)}\put(45,65){$\alpha=0.5$}\end{overpic}&
\end{tabular}
    \caption{The long-time average of the magnetization $\overline{m}_x$ versus the transverse field $h$ for different values of the range of the 
    interaction: $\alpha=0.1$ [panel (a)] and $\alpha=0.5$ [panel (b)]. The scaling collapse is shown in panels (c) and (d) for 
    $\alpha=0.1,\, 0.5$ respectively. As in the previous figures, $T=200$.}
  \label{pd:fig}
\end{figure*}

{We obtain the scaling collapse shown in Fig.~\ref{plotmm_res:fig} {(see Appendix~\ref{exponents} for the details of the scaling procedure)}. } In accordance with the exact solution, the 
DTWA  reproduces the logarithmic corrections (see Fig.~\ref{plotmm_res:fig}). Furthermore, we get a scaling exponent $\delta = 0.47\pm 0.04$ in good 
agreement with the exact exponent $\delta=0.5$ and a critical field $h_c = 1$ corresponding to the exact result. The data collapse, shown in Fig.~\ref{plotmm_res:fig}(b), 
is excellent. For comparison the same scaling is 
shown for the exact diagonalization in Fig.~\ref{plotmm_res:fig}(c). 

\begin{figure}
    \centering
    \begin{tabular}{ccc}
    \begin{overpic}[width=58mm]{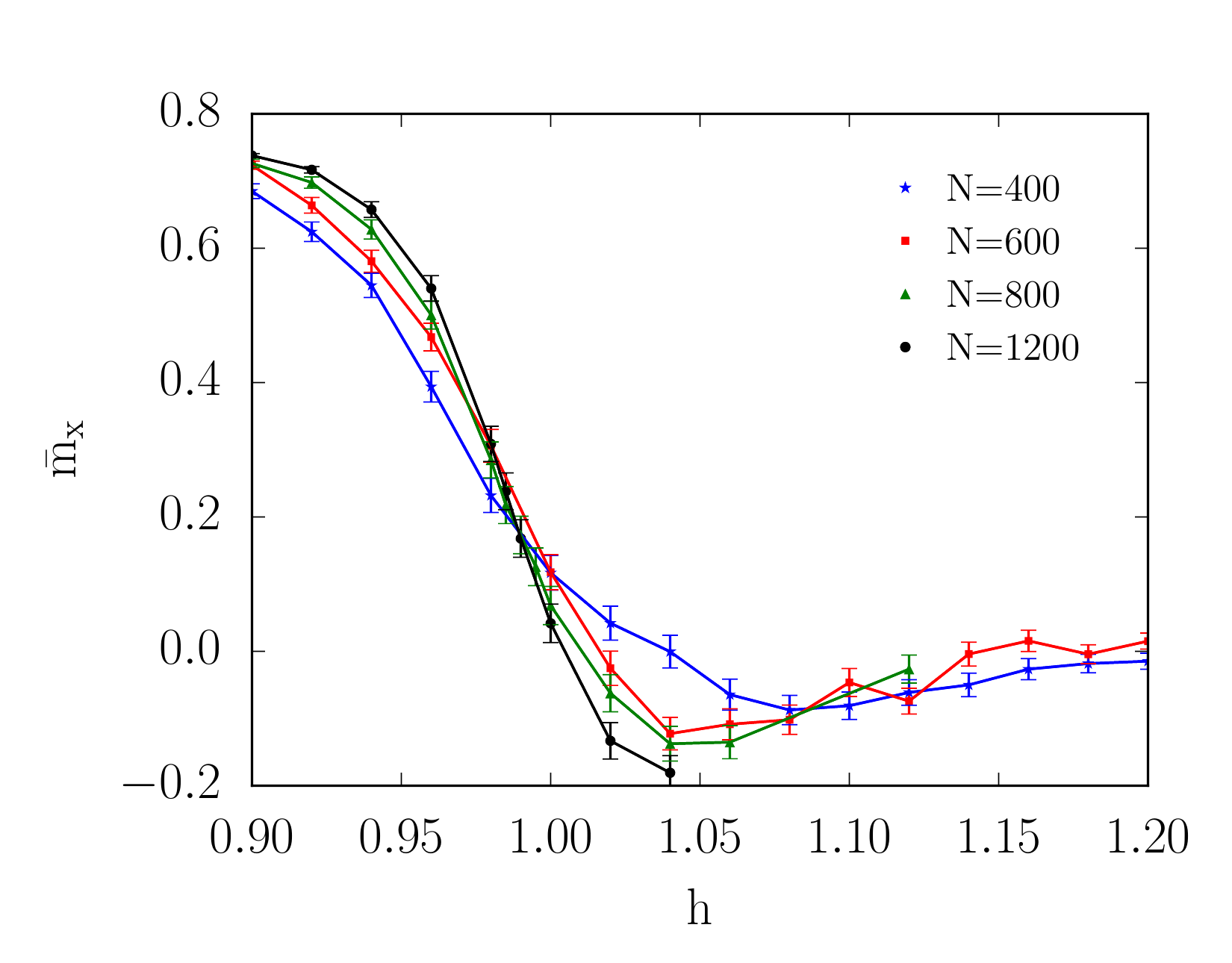}\put(-1,69){(a)}\end{overpic}\\
    \begin{overpic}[width=58mm]{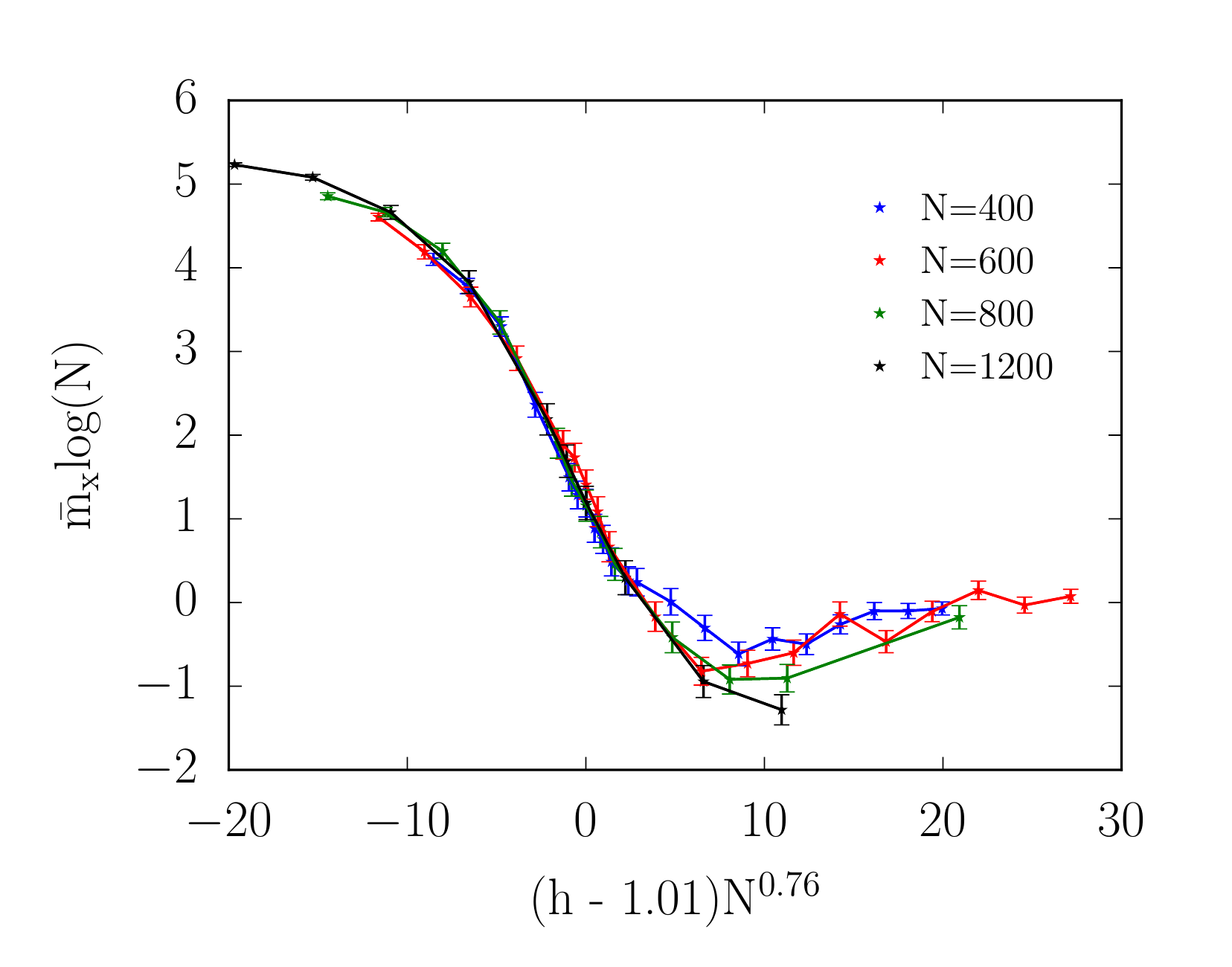}\put(-1,69){(b)}\end{overpic}
    \end{tabular}
    \caption{Scaling plots of the long-time average of the magnetization $\overline{m}_x$ versus the transverse field $h$ at  $\alpha=1.0$ for different values 
    of $N$. The crossing point is obviously close to $h=1$ [panel (a)] but the collapse is not as good as before [panel (b)]. Not withstanding this limitation in the accuracy 
    of the scaling analysis, the exponents are clearly different from the mean-field values. As in the previous figures, also here $T=200$.}
    \label{pd_a1:fig}
\end{figure}

In the following of this section we are going to apply the methods illustrated here to the case with $\alpha\neq 0$.

We first focus on the values of $\alpha \le 1$. We show some examples of $\overline{m}_{x}$ versus $h$ for different sizes $N$ and different $\alpha$ in 
Fig.~\ref{pd:fig} (also in this case the data shown are obtained for $T=200$ where the  observables have already attained their stationary value).
Before doing the finite-size scaling, let us discuss more qualitatively what happens. For $\alpha=0.1$ [Fig.~\ref{pd:fig} (a,c)] and $\alpha=0.5$ 
[Fig.~\ref{pd:fig}(b,d)] we observe a behaviour very similar to the case $\alpha=0$ shown in  Fig.~\ref{plotmm:fig}. In both cases the curves 
show a crossing at  $h_c \sim 1$, the mean-field value. The  tiny deviations from the mean-field are not relevant, only due to the fitting procedure. 
Indeed  we can perform a finite-size scaling with the same method used  for $\alpha=0$ and with the same scaling function as in Eq.~\eqref{scalingb:eqn}. 
In the same Fig.~\ref{pd:fig} (lower panel) we show the collapsed curves. For $\alpha \lesssim 0.5$ the critical behaviour is mean-field like. In particular,
for $\alpha=0.1$ we find $\delta=0.49\pm 0.024$ and for $\alpha=0.5$ we find $\delta = 0.46\pm 0.032$.

A different behaviour is observed at larger $\alpha$ (shorter-range interactions). We show the data for $\alpha=1$ in Fig.~\ref{pd_a1:fig}. The crossing point 
is clearly visible albeit the quality of the data collapse is not as good as in the previous cases. Several points are worth to be discussed. First of all the crossing 
field is still very close to one. The exponent $\delta = 0.76\pm 0.042$, however, deviates significantly from the mean-field value. Although the DTWA does not allow 
ascertain how sizeable is the deviation from the exact scaling analysis, one can be confident in stating that for these parameters $\alpha$ there is still a transition point but 
the critical behaviour deviates from the mean-field. 

Another feature that is worth noticing is that there is a  range of transverse fields (in the disordered region above the critical field) where the magnetization 
becomes negative. Our analysis cannot exclude that this "reentrant" behaviour might still be a feature of the DTWA approximation, not present in more accurate 
analysis. It is however to be noted that this overshooting of the magnetization might be reminiscent of the chaotic behaviour observed 
in the mean-field dynamics of this model~\cite{PhysRevB.100.180402}.

\begin{figure}[b]
  \begin{tabular}{ccc}
  \begin{overpic}[width=60mm]{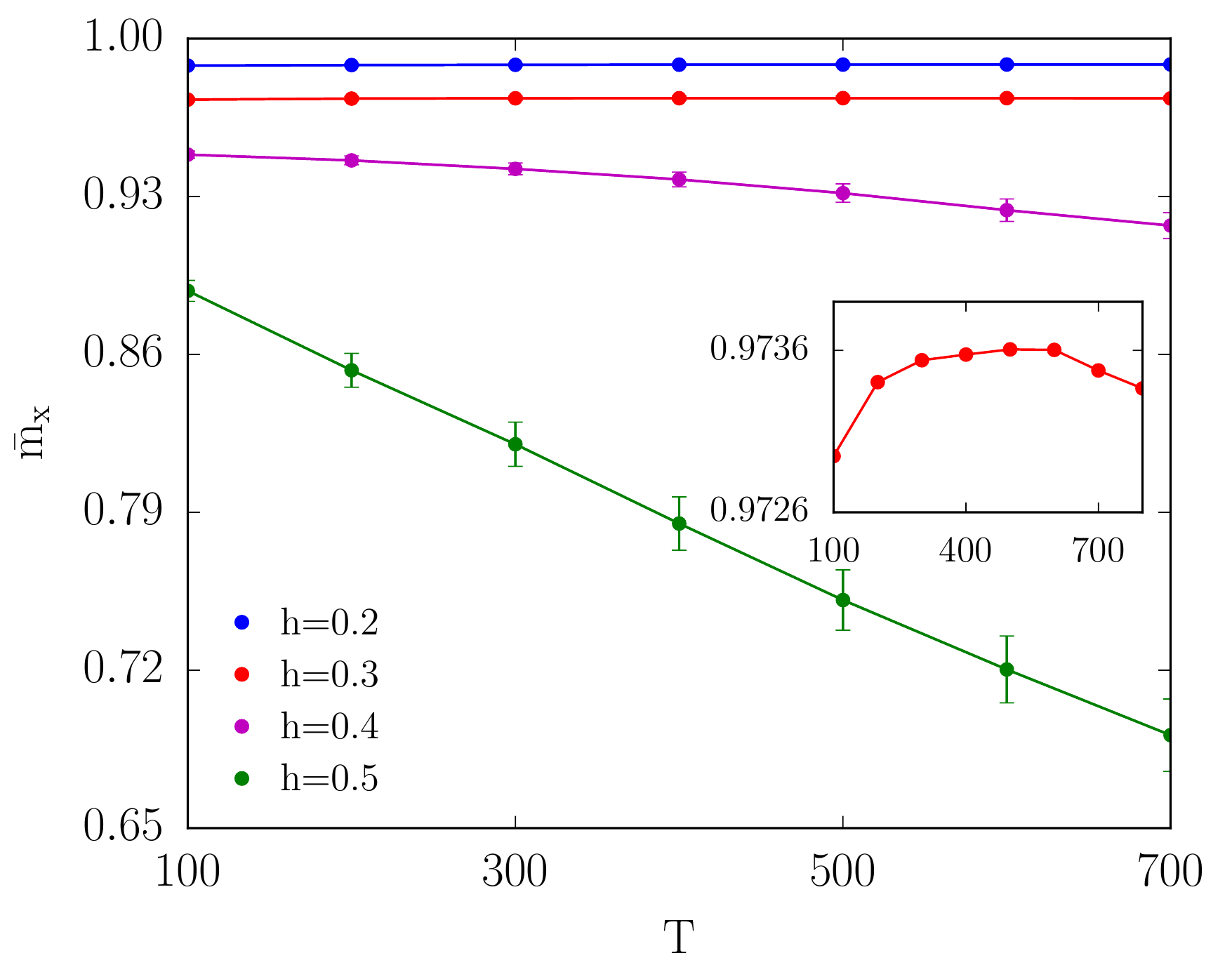}\put(-4,77){(a)}\end{overpic} \\
    \begin{overpic}[width=60mm]{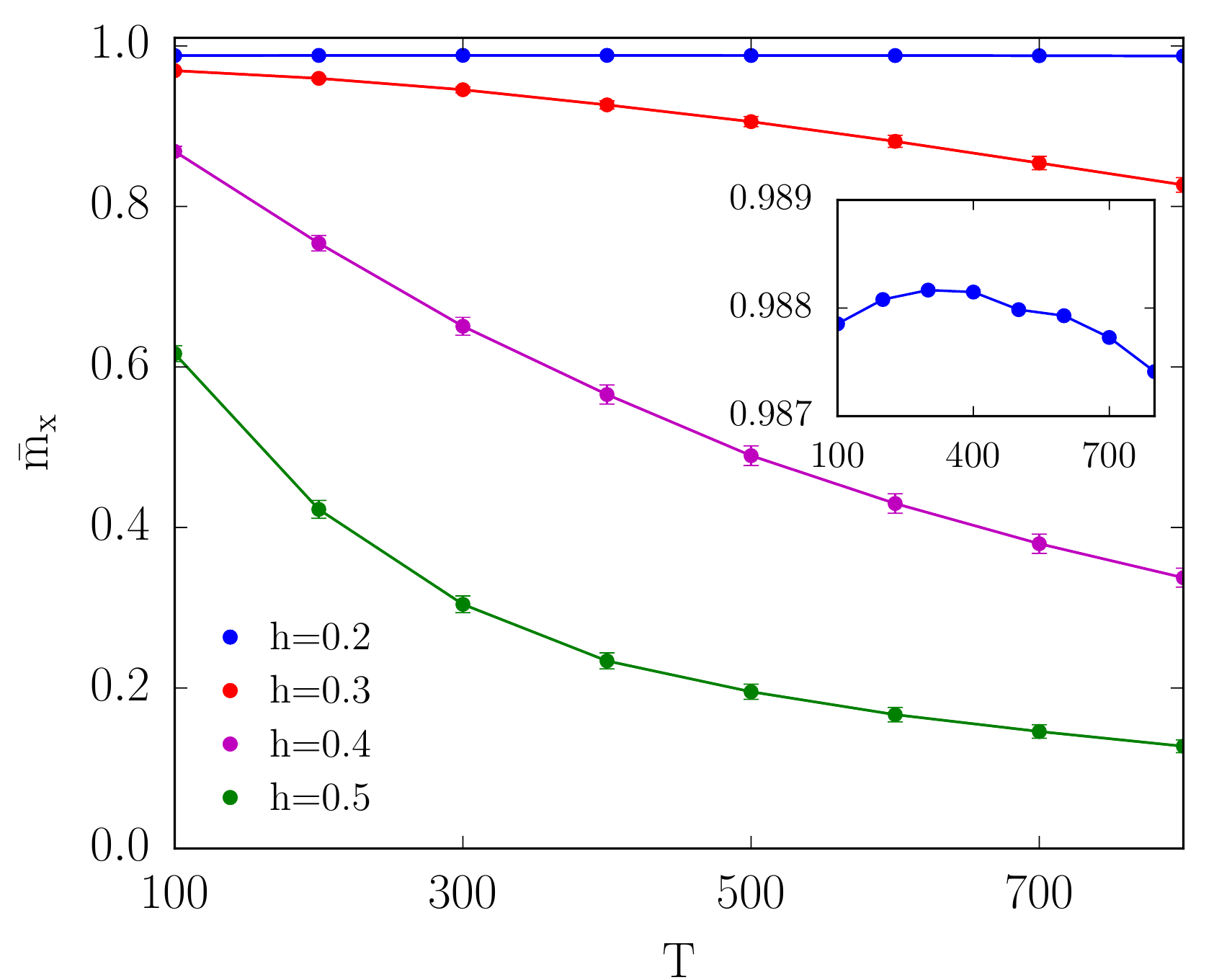}\put(-4,77){(b)}\end{overpic}&
 \end{tabular}
    \caption{{The long-time average of the magnetization $\overline{m}_x$ versus averaging time $T$, [panel (a)] and $\alpha=2$ [panel (b)] $\alpha=3$.  
    The curves show that the time-averaged magnetization decreases with $T$. This behaviour becomes less visible on decreasing $h$ almost disappearing for small $h$. 
    In the insets, a zoom of the curves at $h=0.3$ [panel (a)] and $h=0.2$ [panel (b)] confirm the same trend. The values $h=0.3$ and $h=0.2$ thus given an upper bound to the possible 
    critical field, as extracted by our analysis.  Numerical parameters: $N=100,\, n_r=304$.}}
  \label{m_vs_T:fig}
\end{figure}

{We conclude the analysis of the one-dimensional model by discussing the case of shorter-range interactions, $\alpha \gtrsim 2$.  In this regime DTWA has no more quantitative agreement
with the TDVP methods, so the results have only qualitative value. In this regime we observe that the time-averaged magnetization $\overline{m}_x(T)$ decreases with the averaging time $T$ and never reaches a plateau. This behaviour can be observed for $h$ large enough. }

{We show these results in Fig.~\ref{m_vs_T:fig}. We consider two prototypical cases,  $\alpha=2$ [panel (a)] 
and $\alpha=3$ [panel (b)] and we show the time-averaged magnetization $\overline{m}_x(T)$ versus $T$ for different values of $h$.
For $h$ sufficiently large we see that the average magnetization decreases with $T$ and does not seem to reach a plateau. 
The corresponding slope of this decrease becomes smaller for smaller values of $h$ and for $h=0.2,\,0.3$ the decrease is almost invisible (see the insets which are included for illustration). At small values of the field it would be necessary go to larger times $T$ in order to see this trend.}

{Remarkably, the DTWA gives the same results we have just described for the case of a one-dimensional model with short range interactions, as we discuss in detail in Appendix~\ref{1dcase_shr}. In that case the model is known to show no long-range order in the excited states~\cite{scalettar}, and $\overline{m}_x$ is doomed to vanish whichever is the value of $h$, as can be shown explicitly using the Jordan-Wigner transformation~\cite{suzuki,pappalardi_JSTAT16}. Our DTWA numerics suggests that the situation is the same also for the long-range model with $\alpha \gtrsim 2$, but for sure this is not a proof and the question is still debated~\cite{Silva,2017Halimeh}. From the numerics, 
obviously, we cannot exclude that a transition point still exists at $h_c \ll 1$.}

While in one dimension the short-range case is trivial (there is no DPT), the picture changes drastically by moving to higher dimensions. In the next Section we 
consider the case of a two-dimensional short-range interacting system as defined in Eq.~(\ref{sr-J}).

\subsection{Two-dimensional short-range model} 
\label{2dcase}

This case is of particular importance for several reasons. First of all, to our knowledge, it has never be considered so far. Moreover, we expect that the transition will 
deviate from the mean-field behaviour. This then leads to the question whether the DTWA is capable to detect the transition and its non-mean field type character. If this is 
the case,  {a very important question to understand is if the system thermalizes and the dynamical transition corresponds to a thermal-equilibrium transition.} 
The discussion below will try to address some of these  points by analysing both the magnetization and the Binder cumulant.

\begin{figure}
\begin{tabular}{c}
 \begin{overpic}[width=80mm]{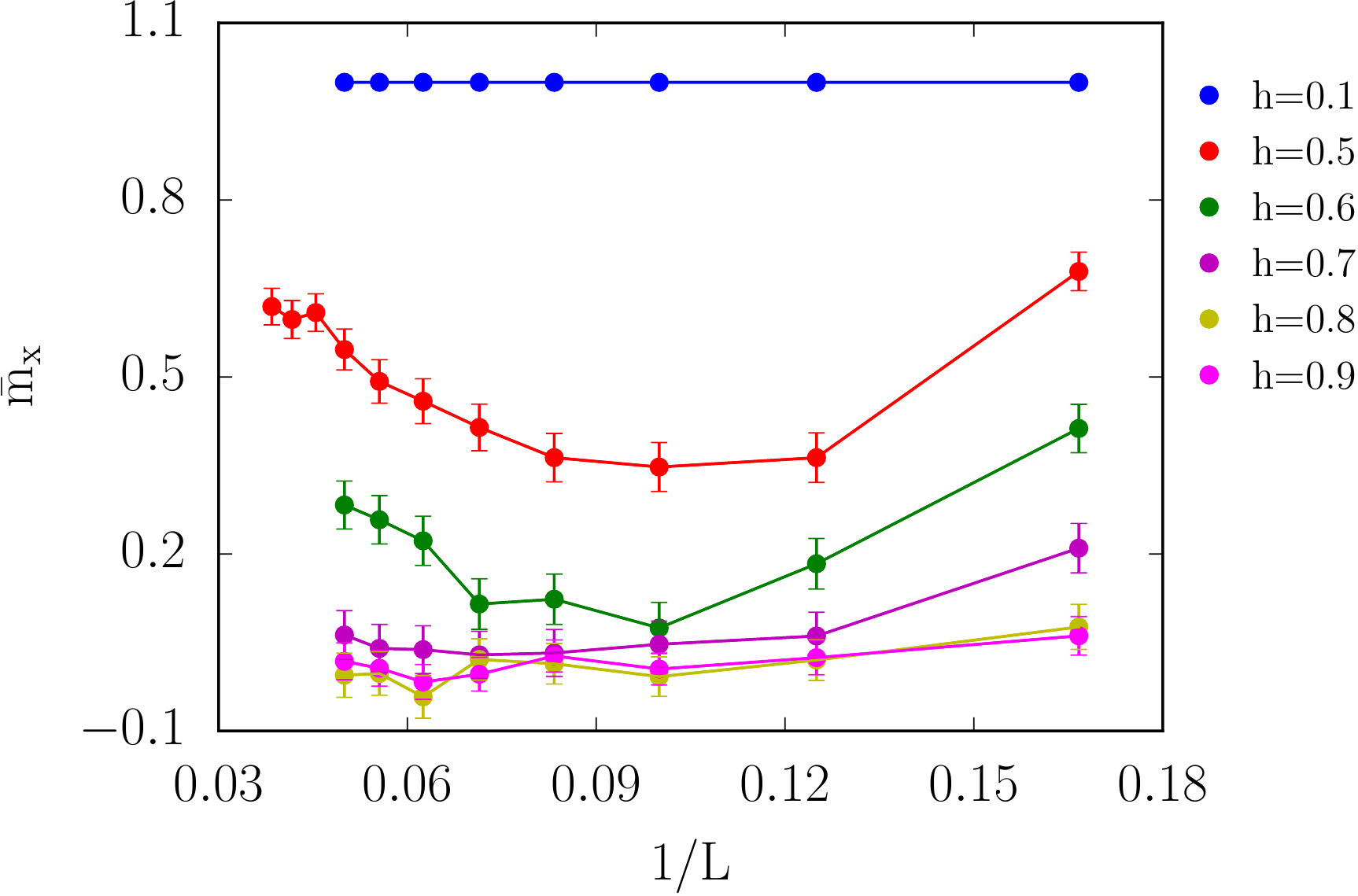}\end{overpic}
\end{tabular}
\caption{{Results of the DTWA for {the long-time average of the magnetization} $\overline{m}_x$ versus $1/L$ in a short-range two-dimensional system for different values of $h$. For $h \gtrsim 0.7$
	the magnetization seems to go to zero in the thermodynamic limit. 
        Numerical parameters: $n_r=1600,\,T=6\cdot 10^5$.}}
\label{mag_vs_L:fig}
\end{figure}

In Fig.~\ref{mag_vs_L:fig} we show the behaviour of the {time-averaged} magnetization as a function of $1/L$ for different values of the transverse field. Here we  take $T=6\cdot 10^{5}$ due to  the long
convergence times (this is essentially the limiting factor that forbids us to consider larger lattice sizes). DTWA indicates the existence of a transition for $h^*\simeq 0.7$.
In the ordered phase the magnetization increases with the system size and tends to converge only for the largest samples.
This type of finite-size effects were observed also in the one-dimensional case where the convergence with size was similarly attained only for $N \sim 100 - 200$.

\begin{figure}
\begin{tabular}{c}
 \includegraphics[width=75mm]{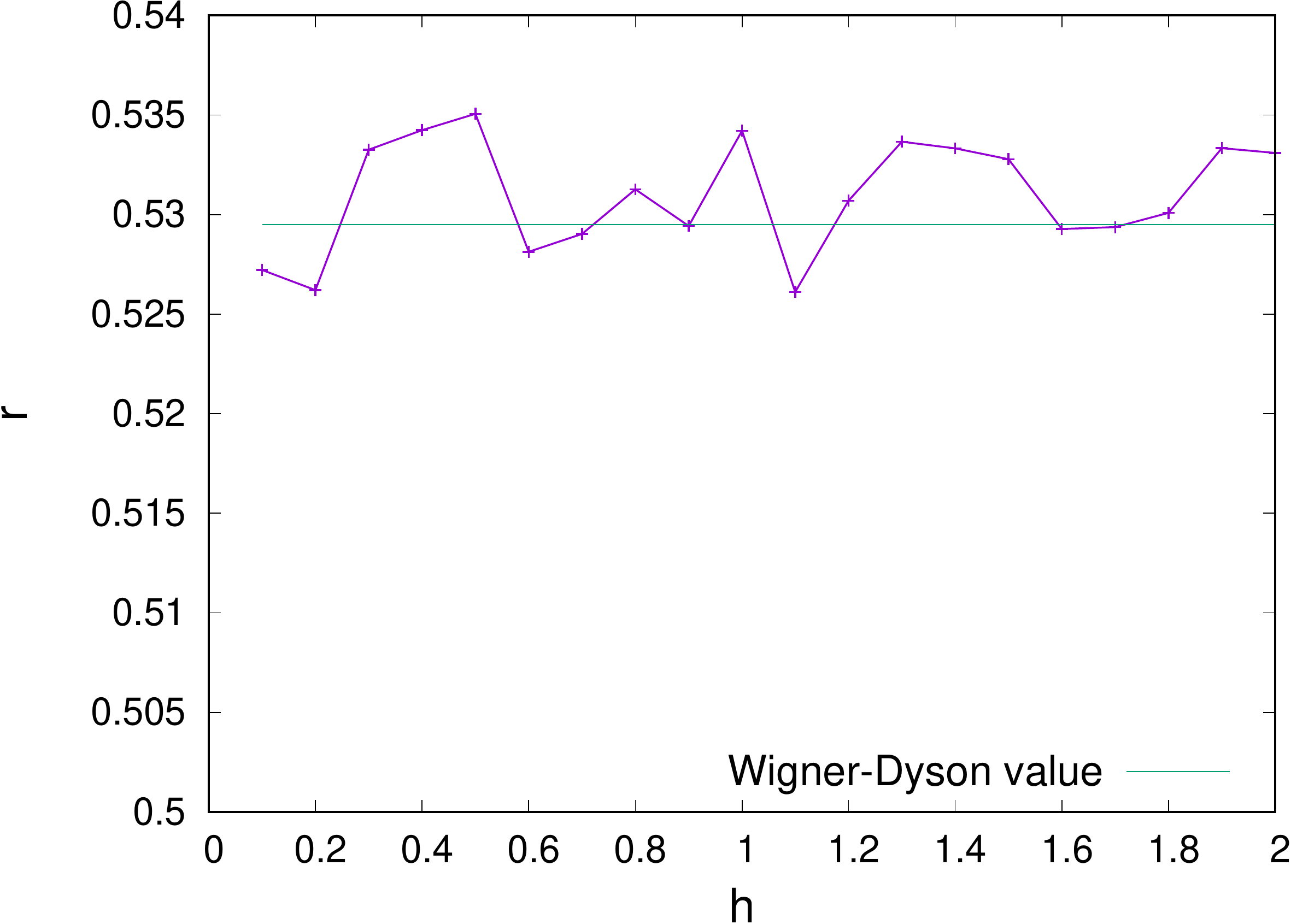}
\end{tabular}
        \caption{level spacing ratio $r$ versus $h$ in the fully symmetric subspace of the two-dimensional model with size $5\times 4$.}
\label{r:fig}
\end{figure}

{We now move to discuss the issue if this transition is the same as the thermal-equilibrium one. First of all we notice that the model is quantum chaotic and thermalizing. We can show the presence of quantum chaos by considering the level spacing distribution and checking that it is near to the Wigner-Dyson 
one~\cite{Haake}. For that purpose we compute the average level spacing ratio $r$ (see~\cite{Oganesyan} for a definition and discussion). Using exact diagonalization in the fully 
symmetric Hilbert subspace of a $5\times4$ model we find a value of $r$ very near to the Wigner-Dyson value $r_{\rm WD}=0.5295$ for all the considered values of $h$ (see Fig.~\ref{r:fig}). 
We therefore expect that the quantum dynamics shows a transition closely corresponding to the thermal one. }


\begin{figure}
\begin{tabular}{c}
 \begin{overpic}[width=80mm]{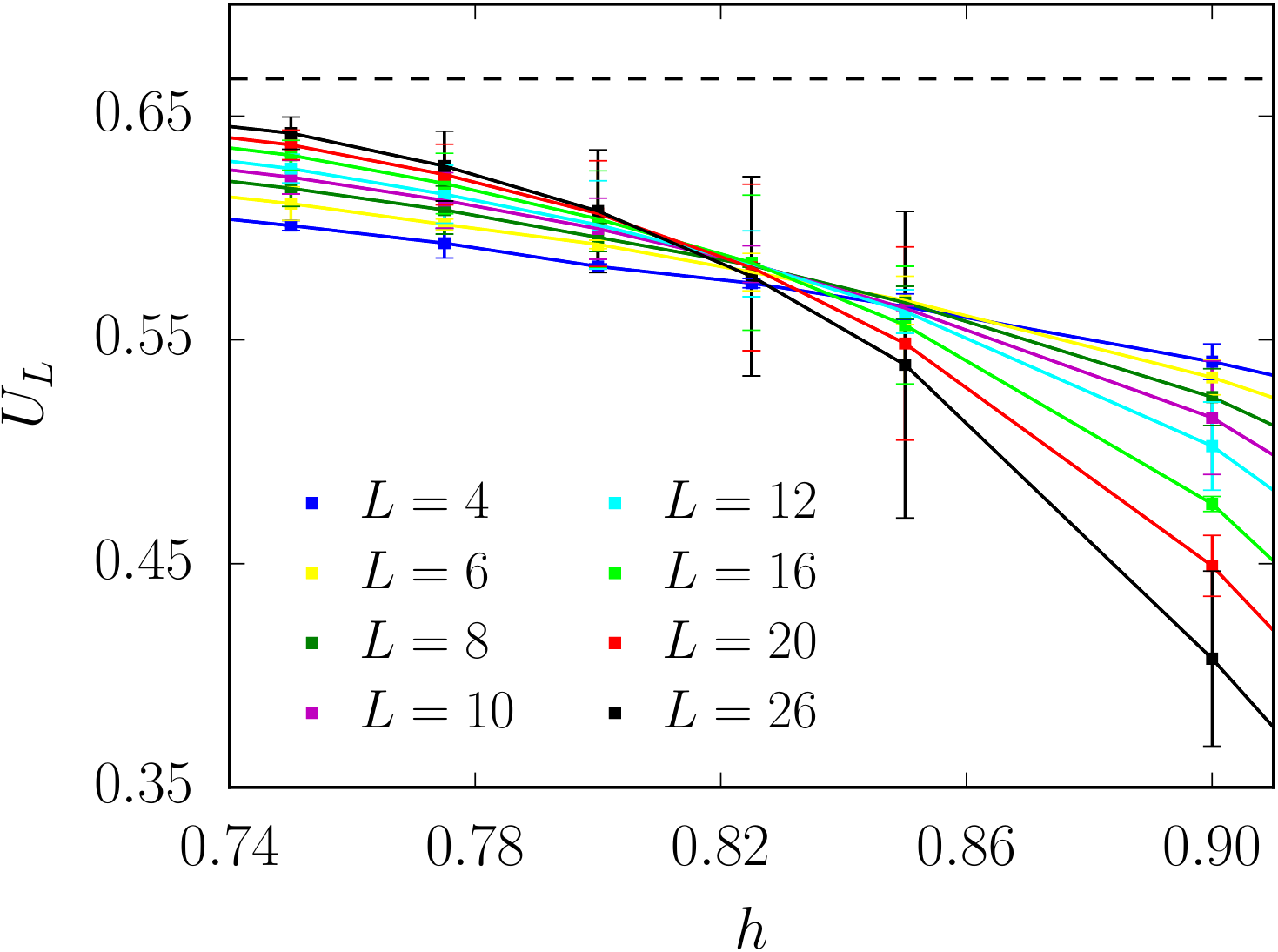}\end{overpic}
\end{tabular}
	\caption{{Binder cumulant at thermal equilibrium obtained via quantum Monte Carlo versus $h$. The considered temperature $\mathcal{T}(h)$ depends on $h$ in such a way that the energy always coincides with the value of the dynamics. Notice the crossing of the curves for different system size at $h_{\rm Th}^*\simeq 0.82$. The error bars indicate a worst-case estimate of the error introduced by estimating the temperature at fixed energy $\mathcal{T}(h)$ from numerical data (not a Monte Carlo error).}}
\label{Bth:fig}
\end{figure}

{We can confirm this expectation by moving to the Binder cumulant analysis. Using this probe, we show that the value of the critical field is not far from the value obtained with quantum Monte Carlo simulations at thermal equilibrium. (We perform the quantum Monte Carlo simulations using the ALPS/looper Library \cite{alps1,alps2,alps3,alps4,alps5}.) In this framework, we take a temperature $\mathcal{T}(h)$ such that the thermal energy coincides with the value of the energy in the DTWA dynamics, and study the properties of the thermal-equilibrium Binder cumulant. It is defined as
\begin{equation}
 U_L(\mathcal{T}(h)) \equiv 1 - \frac{ \mean{\left[\sum_{i=1}^{N}\hat{\sigma}_i^x\right]^4}_{\mathcal{T}(h)} }{ 3\mean{\left[\sum_{i=1}^{N}\hat{\sigma}_i^x\right]^2}_{\mathcal{T}(h)}^2 }\,.
\end{equation}
where $\mean{\cdots}_{\mathcal{T}(h)}$ is the thermal-equilibrium average at the temperature $\mathcal{T}(h)$ defined above. We plot $U_L(\mathcal{T}(h))$ versus $h$ for different values of $L$ in Fig.~\ref{Bth:fig}. We see that the curves for different system sizes cross each other at
$h_{\rm Th}^*\simeq 0.82$. This finding suggests that there is a transition from an ordered to a disordered phase at this value of $h$ (see the general discussion of~\cite{binder}).
The value of $h_{\rm Th}^*$ is not far from the one we have found studying the magnetization with DTWA, suggesting that this model thermalizes and DTWA can catch up to some extent this aspect of the dynamics.} 
%
%
%
%

\begin{figure}[t]
\begin{tabular}{c}
    \begin{overpic}[width=70mm]{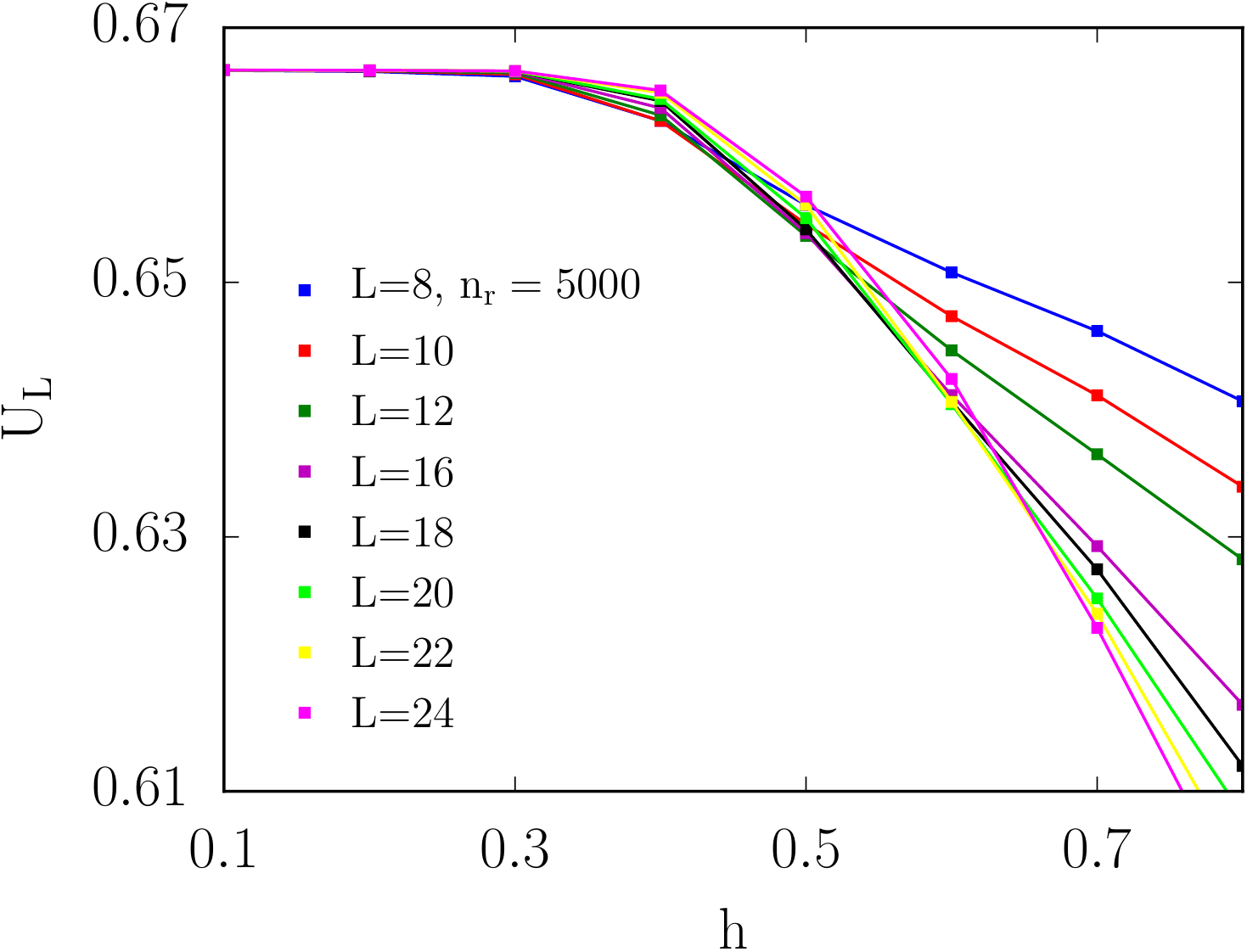}\end{overpic}  
\end{tabular}
\caption{ {The Binder cumulant $U_L$ [Eq.~\eqref{bionder:eqn}] versus $h$ in a short range 2d system for different 
	values of $h$. Numerical parameters: $T=10^4$.}}
\label{bindervsh:fig}
\end{figure}

\begin{figure}
\begin{tabular}{cc}
    \begin{overpic}[width=70mm]{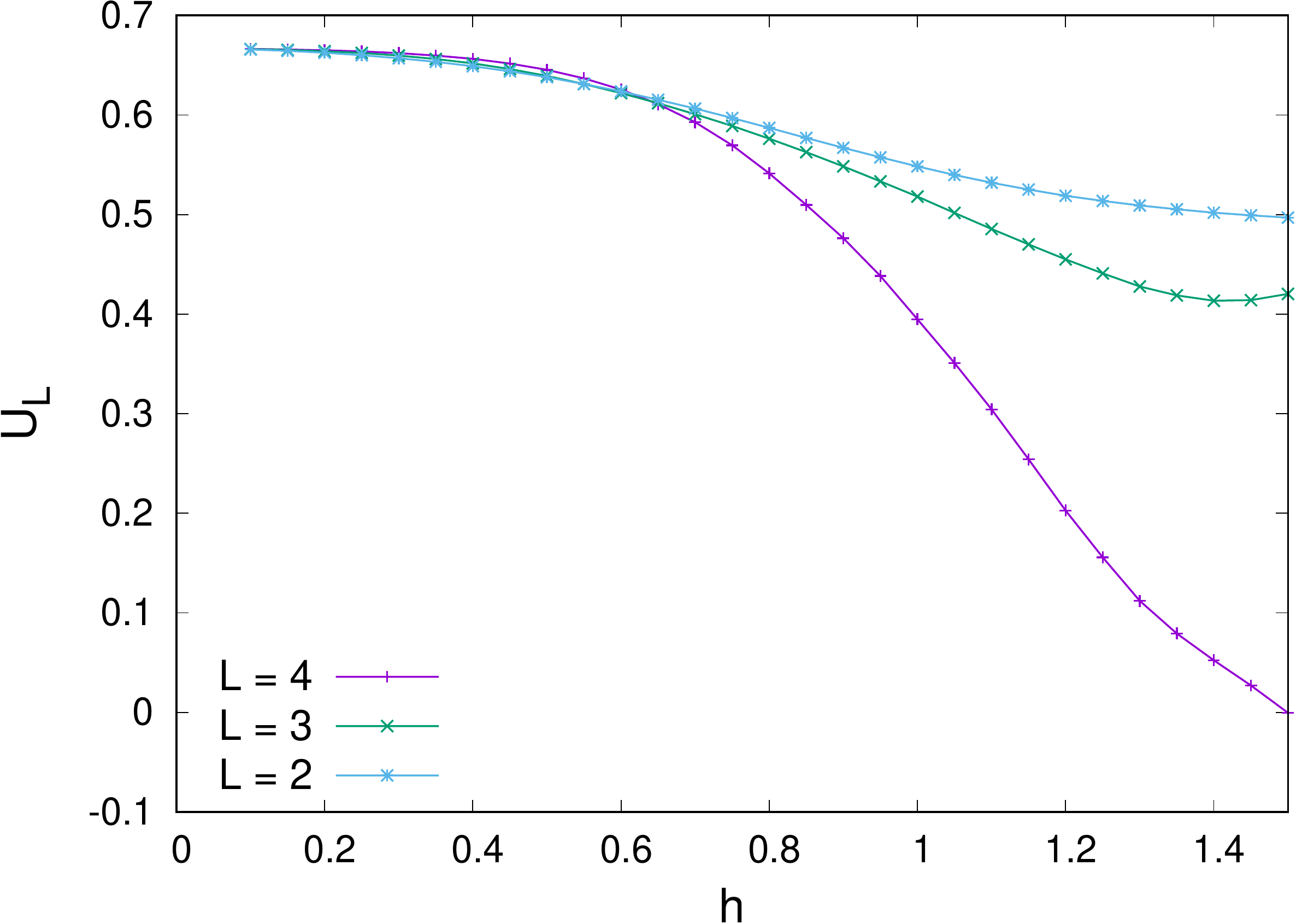}\end{overpic}\\
\end{tabular}
\caption{{Exact diagonalization result of {the Binder cumulant} $U_L$ [Eq.~\eqref{bionder:eqn}] versus $h$ in a short-range 2d system for 
	different values of $L$. 
	Numerical parameters: $T=1000$.}}
\label{bionder:fig}
\end{figure}

{We find further confirmation of these findings by analysing with different numerical methods the time-averaged Binder cumulant $U_L$, defined in Eq.~\eqref{bionder:eqn}. 
We study the dynamics with DTWA and exact diagonalization and we consider the behaviour of $U_L$ versus $h$ for different system sizes.
We show
data for DTWA in  Fig.~\ref{bindervsh:fig} and the ones for exact diagonalization in Fig.~\ref{bionder:fig}. 
Let us first focus on thr DTWA curves in Fig.~\ref{bindervsh:fig}.} 
{The crossing between curves at system sizes $L$ and $L+2$ depends on $L$. For the largest sizes we can numerically attain ($L=22$), the crossing occurs at $h^*\sim 0.65$.
For fields beyond the crossing point,} the Binder cumulant rapidly decreases with $L$.
This is physically sound: The total magnetization is the sum of the local magnetizations which behave as uncorrelated 
random variables at large $h$ because the correlation length is very short. The sum of uncorrelated random variables tends to a Gaussian as the number of random variables 
increases and for a Gaussian the Binder cumulant vanishes. For small values of $h$, on the opposite, $U_L$ increases with $L$. Therefore a crossing 
point between curves for different $L$ appears. 

\begin{figure}[t]
\begin{tabular}{c}
    \begin{overpic}[width=70mm]{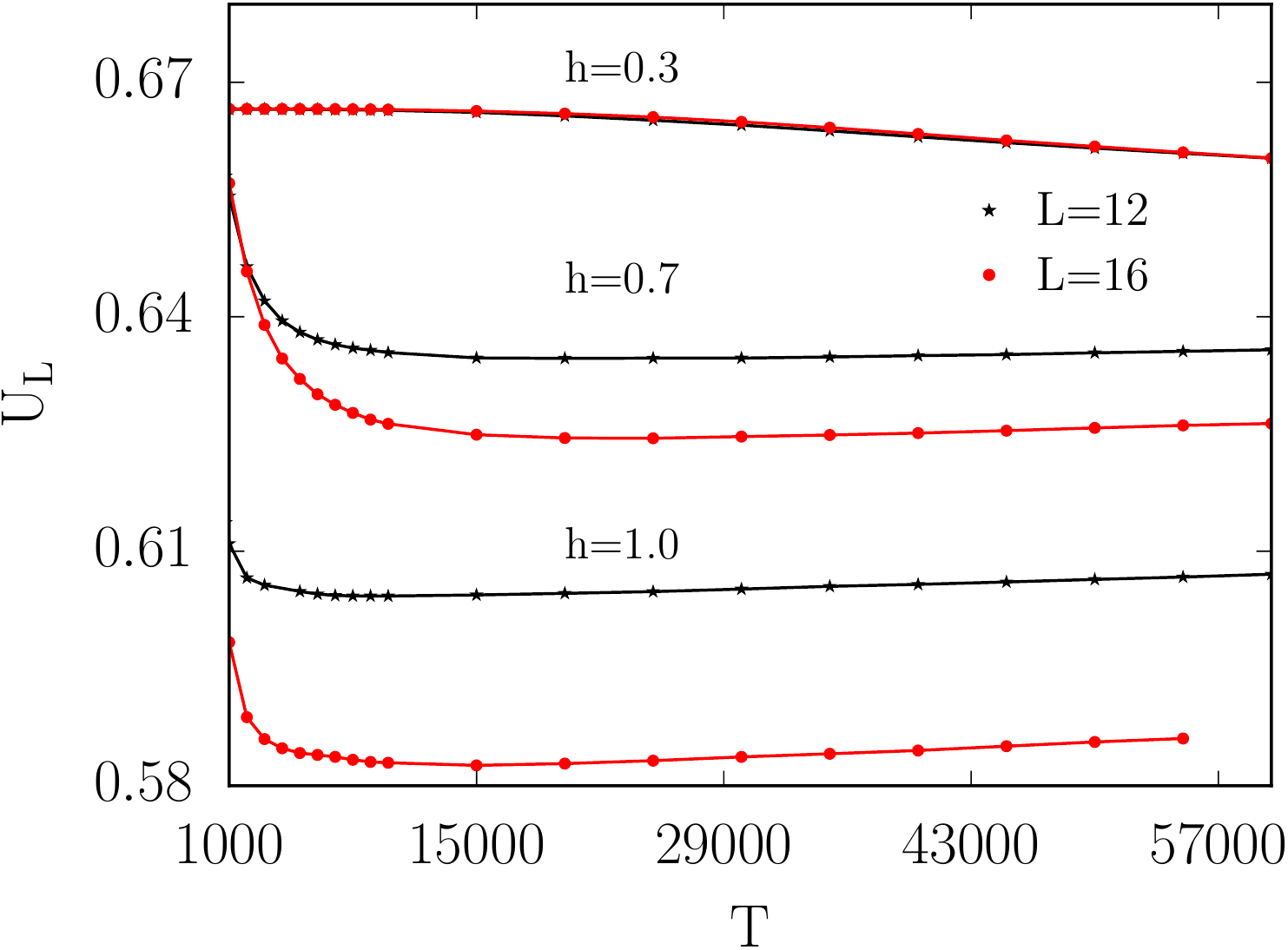}\end{overpic}  
\end{tabular}
\caption{ {The Binder cumulant $U_L$ [Eq.~\eqref{bionder:eqn}] versus $T$ in a short range 2d system for different 
	values of $h$ and $L$. Notice the metastable plateau. Numerical parameters: $T=10^4$.}}
\label{bindervsT:fig}
\end{figure}

The Binder cumulant has been evaluated averaging over a time ($T=10^4$) shorter than the time needed to attain an asymptotic value in the DTWA scheme. The point is that, before this asymptotic value, the Binder cumulant attains a metastable plateau {in the DTWA scheme}: We show some examples in Fig.~\ref{bindervsT:fig}. This plateau gives rise to the crossing behaviour we can see in Fig.~\ref{bindervsh:fig} while the asymptotic value does not. The metastable plateau therefore shows a behaviour more similar to the ones given by quantum Monte Carlo (Fig.~\ref{Bth:fig}) and by exact diagonalization (Fig.~\ref{bionder:fig}). This suggests that in this context DTWA gives physically more sound results for a finite time, although the difference between the metastable plateau and the asymptotic value is very small. {We remark that this plateau is an effect of the approximation and does not correspond to any prethermalization behaviour in the actual physics}


{We study the behaviour of the Binder cumulant also by means of exact diagonalization.  
In Fig.~\ref{bionder:fig} we show the exact-diagonalization Binder cumulant versus $h$ for small system sizes. The trend is the same as that observed in Fig.~\ref{bindervsh:fig}.
The crossing occurs around $h^*\sim 0.6$, which is in good agreement with the value found using DTWA.
We stress again that for increasing system size the Binder cumulant tends to $2/3$ in the ordered phase and to 0 in the disordered one, exactly as it occurs in the thermal-equilibrium case.}

In conclusion, for the two-dimensional short-range case there is a dynamical transition closely corresponding to the thermal one due to the fact that 
the system appears quantum chaotic and thermalizing. Remarkably, DTWA can see the existence of this transition.

\section{Conclusions} 
\label{sec5}
In conclusion we have used DTWA to study the dynamical quantum phase transition in Ising spin models. Our aim was exploring the existence of a transition between an 
ordered and a disordered phase in the steady state and the properties of this transition focusing on a local order parameter, the time-averaged longitudinal magnetization. 

We have first focused on the long-range one-dimensional case where interactions decay with the power $\alpha$ of the distance. Here we have compared DTWA with 
numerically exact results (exact diagonalization for $\alpha=0$ and TDVP) and we have found a good agreement. Thanks to the good scalability of DTWA, we have 
done a finite-size scaling of the time-averaged longitudinal magnetization and we have studied the critical exponents of the transition between ordered and disordered 
phase. For $\alpha$ small ($\alpha=0.1,\,0.5$) we have found the same critical exponents as the mean-field case ($\alpha=0$). For $\alpha=1$ we have found critical 
exponents significantly different from the mean-field case and we have found that the magnetization changes sign in the critical region. We do not know if this is a 
physical result or an effect of the DTWA approximation which should not work very well in the critical region due to the long-range correlations of the physical system. 
{For $\alpha \gtrsim 2$  we have} found no scaling at all with the system size and we have put a lower bound to the value of $h$ for which the longitudinal 
magnetization vanishes at long times. We argue that this is most probably the case also for smaller $h$ but we cannot see it due to the extremely long convergence 
times in the DTWA scheme (this is the same situation occurring if we apply DTWA to a short range one-dimensional Ising model). 

{We further considered the 2d short-range model, not considered in this context so far, again applying the DTWA approximation. Our data confirms that the 
DTWA is able to capture the existence of a transition and the value of the critical field compares well with the one of a corresponding thermal transition. 
We argued that this is physically sound showing that the model is quantum chaotic by means of exact diagonalization. In order to attempt a scaling analysis and thus 
to confirm that the associated critical exponents are the thermal ones it would be necessary to consider even larger system sizes, which might be an interesting prospect for the future.}

{Our work can also be considered as a contribution towards the clarification of the range and the limitations of
qualitative and quantitative applicability of the DTWA}

%

\section{Acknowledgments}
We would like to acknowledge fruitful discussions with Silvia Pappalardi. R.~F. acknowledges financial support from the Google Quantum Research Award. R.~K. acknowledges financial support from the ICTP STEP program.
M.~S. was supported through the Leopoldina Fellowship Programme of the German National Academy of Sciences Leopoldina (LPDS 2018-07) with additional support from the Simons Foundation.
This project has received funding from the European Research Council (ERC) under the European Unions Horizon 2020 research and innovation programme (grant agreement No. 853443), and M.~H. further acknowledges support by the Deutsche Forschungsgemeinschaft via the Gottfried Wilhelm Leibniz Prize program.

\section{Appendices}

\appendix
\section{DTWA sampling}
\label{samplingAPP}
%

{
As discussed in Sec.~\ref{DTWA:sec}, in the DTWA approach one has to solve classical equations of motions for different random initial configuration. 
Physical quantities are obtained upon averaging over  this initial distribution. In this Appendix we report on some details of the sampling procedure 
we used to obtain the results reported in the body of the paper.

First of all it is important to understand how the results depend on the number of random initial realizations $n_r$. In Fig.~\ref{fig:order_nr} 
we consider the dependence of the average magnetization as a function of the number of initializations $n_r$. We show the case of $\alpha= 0.1$; the 
behaviour is however quite generic. 
Away from the critical field $h_c$, the order parameter $\overline{m}_x$ converges very rapidly to its asymptotic value and no significant changes happen 
by increasing $n_r$. Since we are interested in determining transition points, the behavior $\overline{m}_x$ as a function of $n_r$ is more notable in the 
critical region.  Close to the transition point the convergence with the number of realizations is slower. In any case  after few hundreds of initial configurations
the results seem stable. We choose $n_r=504$ for most of the calculations, if not stated otherwise.

\begin{figure}[b]
	\centering
		\begin{overpic}[width=81mm]{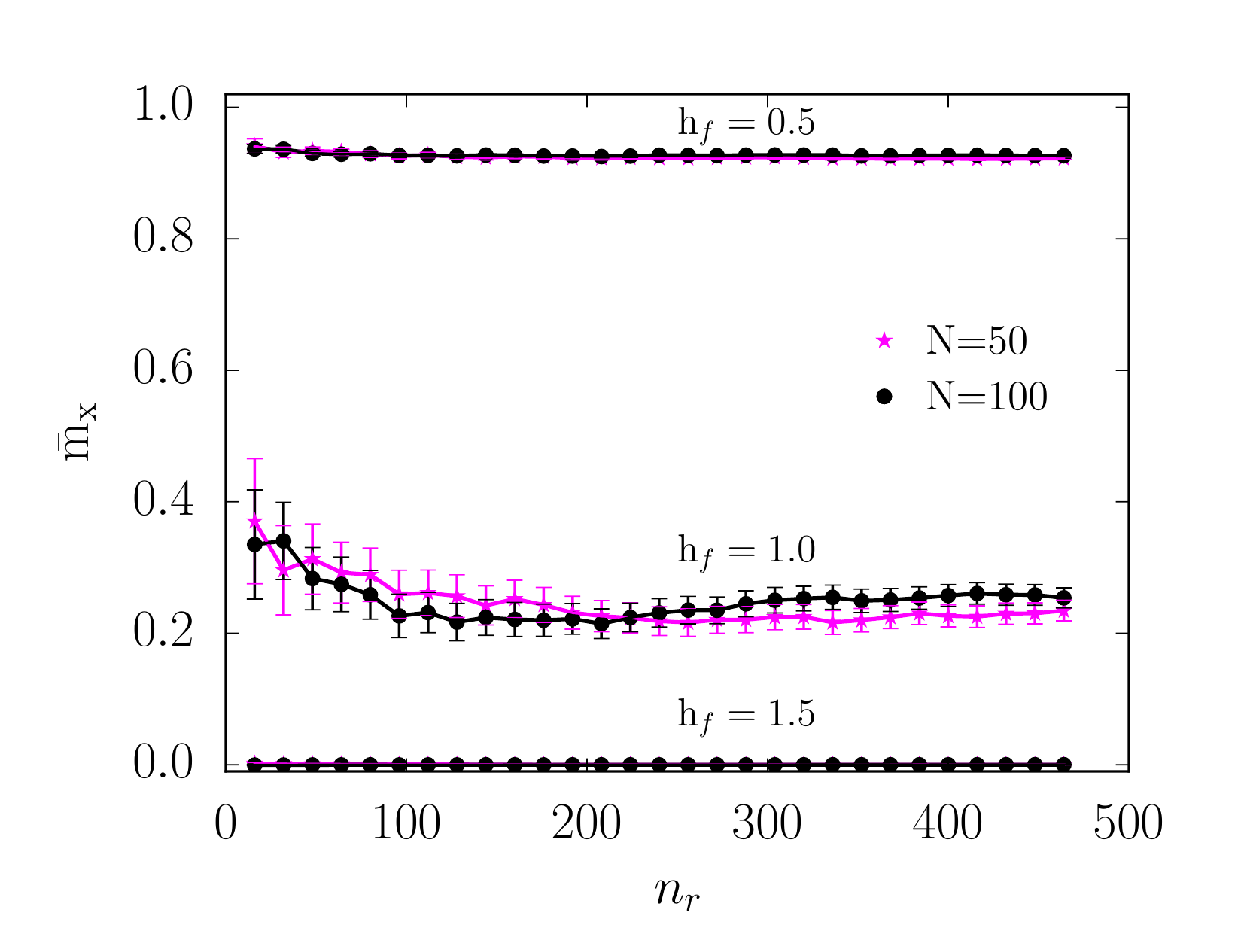}\end{overpic}
	\caption{{{The long-time average of the magnetization} $\overline{m_x}$ versus number of realizations $n_r$ for three different values of the transverse field. 
	The convergence changes depending on the distance from the critical point. However, in all the shown cases, averaging over 100 - 200 configurations 
	already guarantees that the obtained result is reliable. In the case shown here  $\alpha = 0.1$. We tested that this behaviour is quite generic.}}
	\label{fig:order_nr}
\end{figure}

%

In addition to the number of initial configurations over which performing the sampling, another aspect to consider is the choice of the sampling scheme.
Indeed, using phase point operator $\hat{A}_\alpha$ one can map each basis state of Hilbert space to a point in phase space. There are different possible choice 
of this phase operator and the one shown in Eq.(\ref{ini:eqn}) is not the only one. Any other possible choice for phase operator can be derived by some unitary 
transformation, $\hat{A}^{\prime}_{\beta}=\hat{U}\hat{A}_{\beta}\hat{U}^{\dagger}$. 

In ~\cite{2018Czischek} the following phase operator was considered (more details about this construction can be found there)

\begin{equation} \label{ini:eqn1}
    \hat{A}^{\prime}_{\beta}=\frac{\boldsymbol{1}+\textbf{s}^{\prime}_{\beta}\cdot\hat{\boldsymbol{\sigma}} }{2}
\end{equation}
\begin{figure}
\begin{tabular}{ccc}
\begin{overpic}[width=60mm]{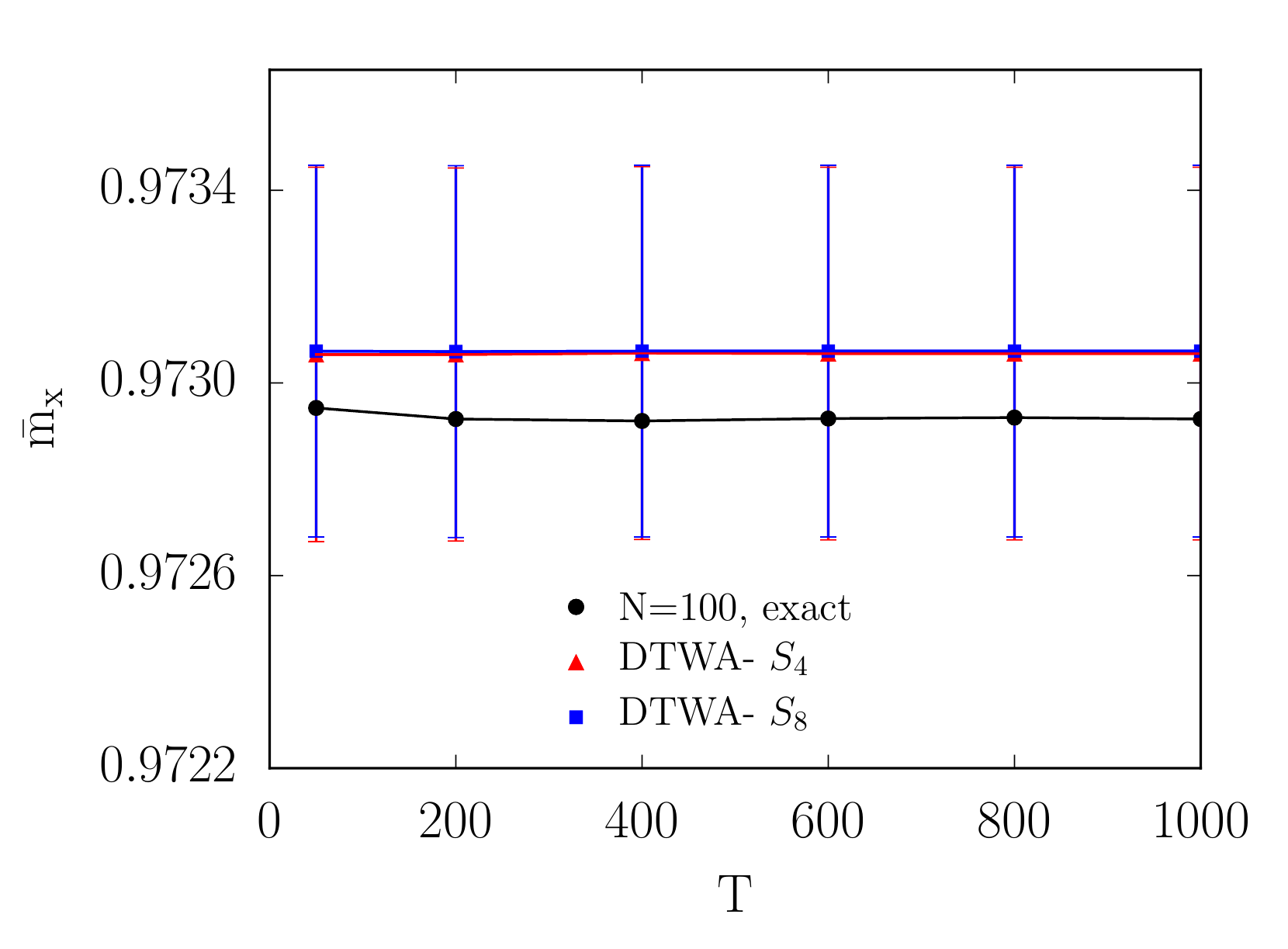}\end{overpic} \\
\begin{overpic}[width=60mm]{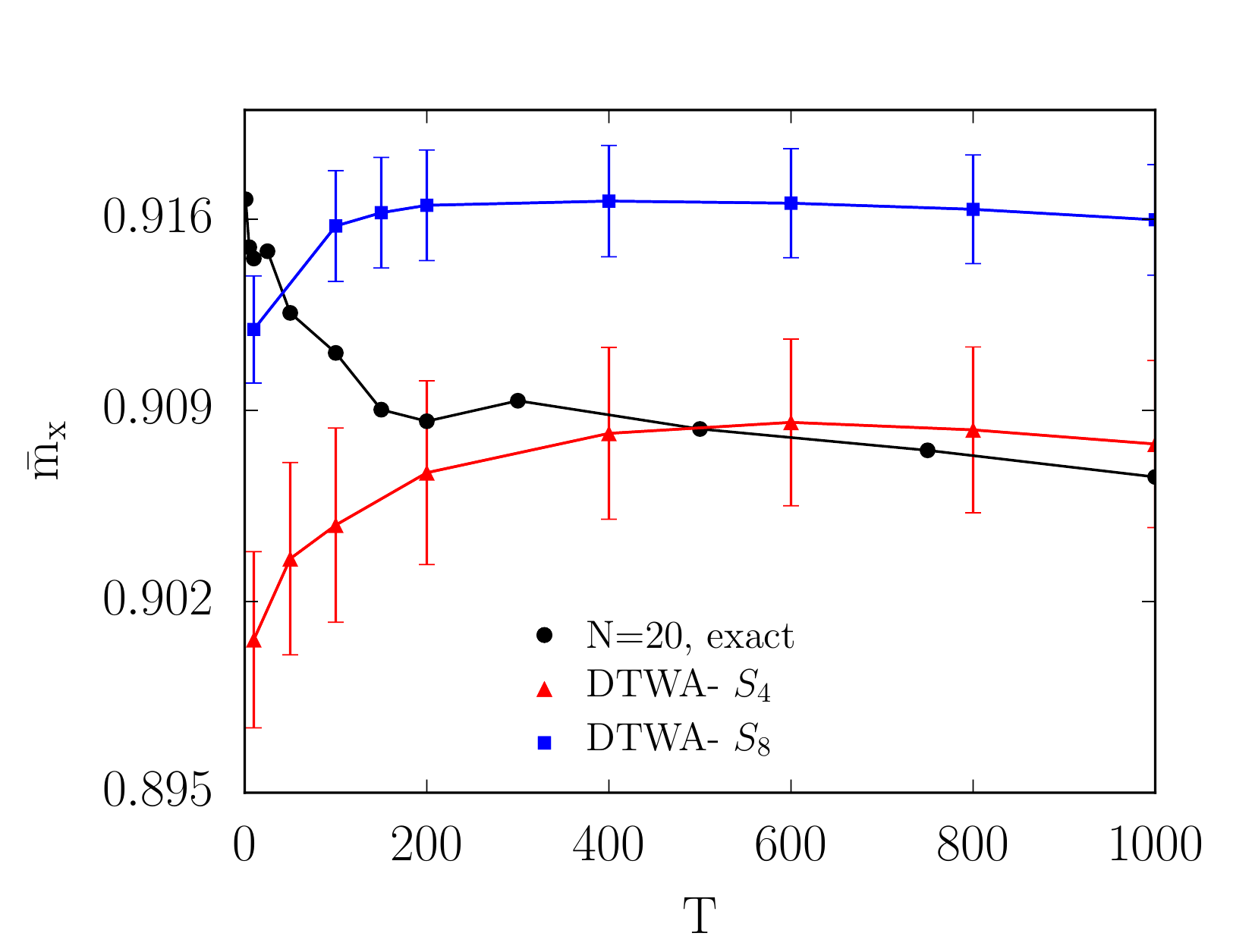}\end{overpic}
\end{tabular}
	\caption{{The time average of the magnetization} $\overline{m}_x$ versus the averaging time $T$: Comparison of the results obtained with exact diagonalization and different sampling schemes of DTWA
	(the sampling schemes $S_4$ and $S_8$ are defined and discussed in~\cite{2018Czischek}). 
	We consider $\alpha=0.0$, $N=100$ (upper panel) and $\alpha=1.0$, $N=20$ (lower panel). Other parameters: $h=0.32$, $n_r=2000$.}
\label{m_vs_T_diff_sam:fig}
\end{figure}
where $\boldsymbol{s}^{\prime}_\beta$ can take the values $\left(\begin{array}{ccc}1&-1&1\end{array}\right)$, $\left(\begin{array}{ccc}-1&-1&
-1\end{array}\right)$, $\left(\begin{array}{ccc}1&1&-1\end{array}\right)$ and $\left(\begin{array}{ccc}-1&1&1\end{array}\right)$ and
$\hat{\boldsymbol{\sigma}}=\left(\begin{array}{ccc}\hat{\sigma}^x&\hat{\sigma}^y&\hat{\sigma}^z\end{array}\right)$ which is obtained by flipping the sign of the 
second component of $\boldsymbol{s}_\beta$. 

Fig.\ref{m_vs_T_diff_sam:fig} show the comparison, as a function of the averaging time $T$ for two different values of $\alpha$. DTWA is further compared to 
exact diagonalization.  In the fully connected case ($\alpha = 0$) the different samplings lead to essentially the same result and agree with the exact 
diagonalization data. Smaller distances are observed in the bottom panel for the case $\alpha=1$. It should be noted that deviations appear only at the the third
decimal digit.  These differences may be important only very close to the transition point and may also contribute to the uncertainties in  the scaling plots that 
we observe for  $\alpha \sim 1$.  However, the analysis of the present work does not depend on the sampling scheme.

\section{Short-range model in one dimension}
\label{1dcase_shr}

In the case of spin chain with short-range interaction there is no ordered non-equilibrium steady-state~\cite{scalettar,suzuki,pappalardi_JSTAT16} (it corresponds to the long-range one-dimensional model studied in the limit of very large
$\alpha$). It is useful to check this result with DTWA as an additional test of its quality. Following the same approach used to argue the absence of 
a critical point for $\alpha \gtrsim 2$ we analyse how the magnetization scales with $T$ for different values of the transverse field. The result of this analysis is 
presented in Fig.~\ref{m_vs_T_shr:fig}. Down to $h=0.1$ the steady state magnetization (at large $T$) tends to zero (top panel). The inset in the top panel shows that in 
order to see the suppression of the magnetization at large $T$ one should go to very large values.  In the top panel we considered a chain of length $N=100$. Because of 
the short-range correlations in this case, the behaviour is essentially independent of $N$ as displayed by the bottom panel of Fig.~\ref{m_vs_T_shr:fig}.

\begin{figure}
\vspace{0.5cm}
\begin{tabular}{ccc}
\begin{overpic}[width=53mm]{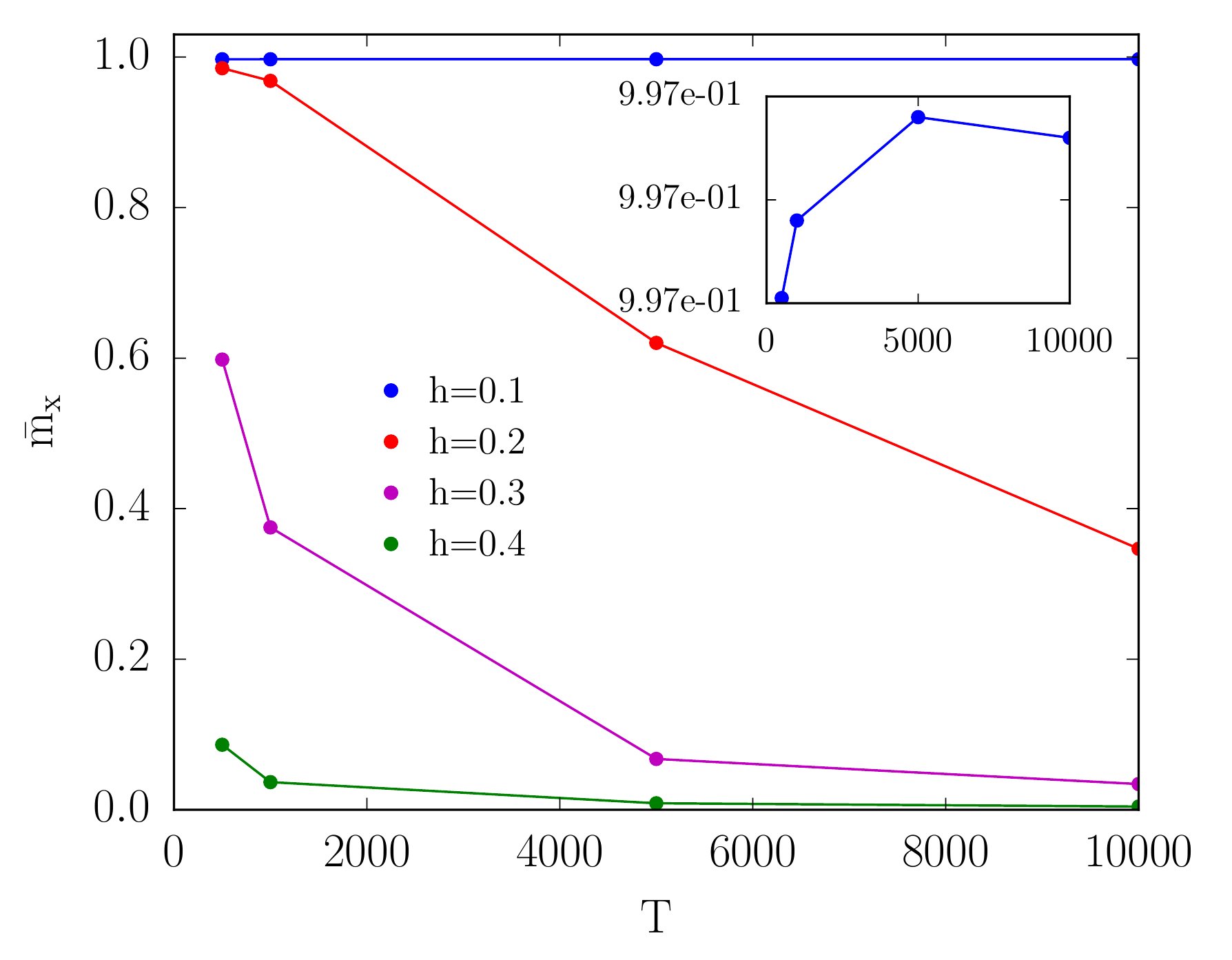}\put(-4,73){(a)}\end{overpic} \\
\begin{overpic}[width=55mm]{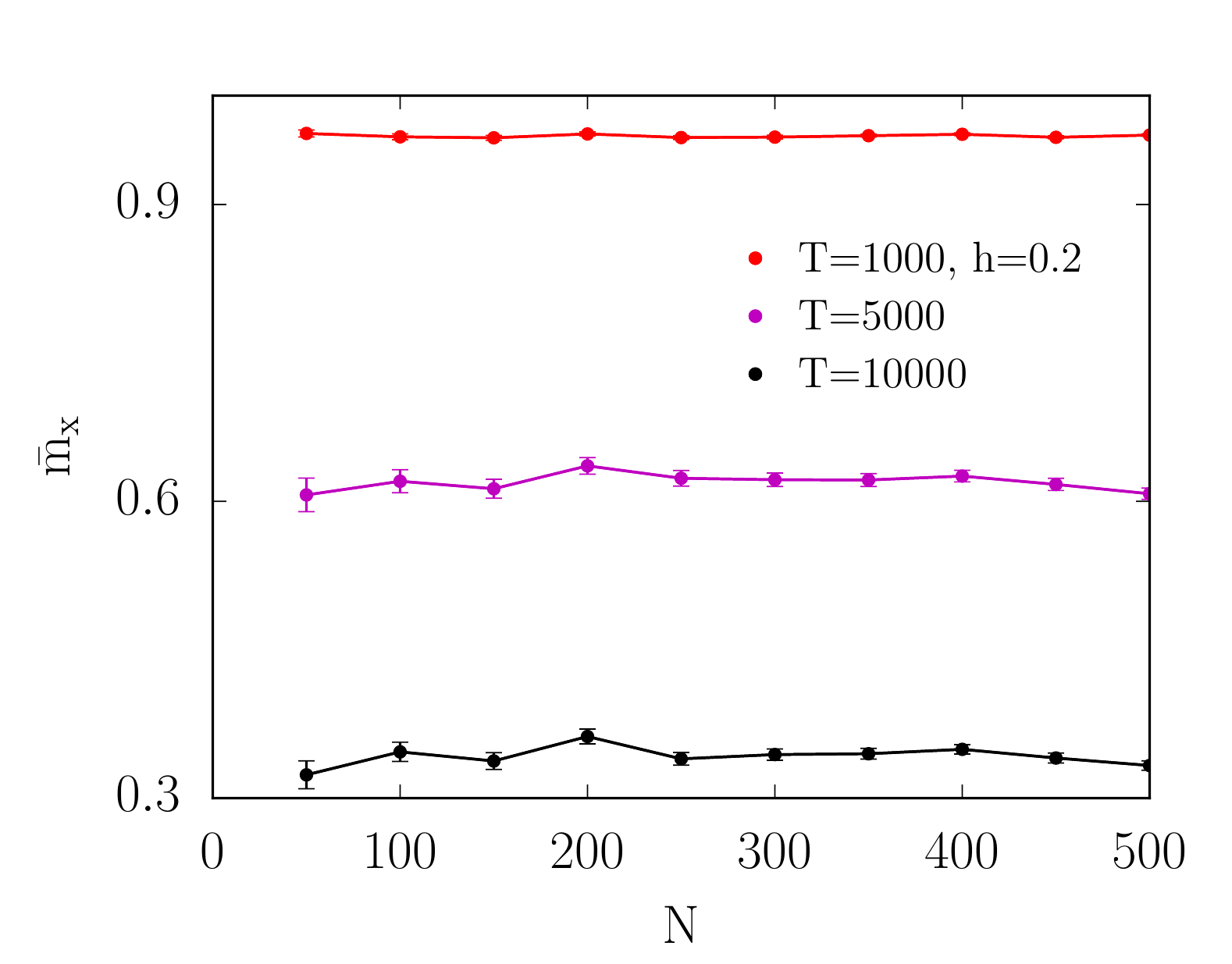}\put(-1,69){(b)}\end{overpic}
\end{tabular}
\caption{(Top panel) {The time-average of the magnetization} $\overline{m}_x(T)$ versus averaging time $T$ {in the Ising chain with short-range interaction. For larger $h$, the 
average magnetization decreases with $T$ towards 0 without ever reaching a plateau: this suggests that $\overline{m}_x(T)\stackrel{\footnotesize T\to\infty}{\to}0$ as in the actual physics. For $h=0.1$ this decay behaviour can only be seen very slightly (inset), but this is an artifact of the DTWA and not a physical effect.} Numerical parameters: $N=100$, 
$n_r=304$. (Bottom panel) Plot of the magnetization versus the system size $N$. The correlations are very short-range for this model and this is reflected in the 
insensitivity on $N$ of the average.}
\label{m_vs_T_shr:fig}
\end{figure}

}

\section{Determination of the critical exponents}
\label{exponents}

{
In order to determine the best approximations to $h_c$ and to the exponent $\beta$, we find the values of $h$ and $\beta$ such that the distance function between the magnetization curves at different $N$
\begin{equation}
d_\beta(h)=\sum_{N,\,N'<N}\abs{N^{\beta}\overline{m}_{x,\,N}(h)-{N'}^{\beta}\overline{m}_{x,\,N'}(h)}
\end{equation}
is minimum. In this way we find $h_c=1.008\pm0.01$, a value very near to the exact one $h_c=1$. Moreover, we find $\beta=0$, as we can see in Fig.~\ref{fig:ditance_exponent}, but we scale the magnetization with $\log N$ in order to take into account the logarithmic corrections. For finding the optimal $\delta$, we minimize with respect to $\delta$ the cost function
\begin{widetext}
	\begin{equation}
	D_\delta = \frac{\sum_{N',\,N<N'}\int\ud x\left[\overline{m}_{x,\,N}(h_c+N^{-\delta}x)-\overline{m}_{x,\,N'}(h_c+{N'}^{-\delta}x)\right]^2}
	{\sum_{N',\,N<N'}\int\ud x\left[\overline{m}_{x,\,N}^2(h_c+N^{-\delta}x)+\overline{m}_{x,\,N'}^2(h_c+{N'}^{-\delta}x)\right]}\,.
	\end{equation}
\end{widetext}

The errorbars in $\delta$ are evaluated in the following way. If we have to perform our minimization procedure on a set of $K$ data curves, we consider all the $K$ distinct subsets of $K-1$ curves. In each of these subsets we perform the minimization procedure and then we get $K$ different values of $\delta$. The standard deviation of these $K$ values of $\delta$ provides the errorbar.

In Fig.~\ref{fig:ditance_exponent} we consider in detail an example of application of our method. In panel (a) we show the minimum distance versus $\beta$, while in Fig.~\ref{fig:ditance_exponent}(b) we show the cost function versus $\delta$ for different $\alpha$ and $h_c$ found using the logarithmic scaling (see below Eq.~\eqref{scalingb:eqn}). In order to perform the integration we apply a cubic spline interpolation. The dependence of $D_\delta$ on $\delta$ is shown in Fig.~\ref{fig:ditance_exponent}(b); we find the minimum in $\delta=0.47$, as we have elucidated in the main text.
%
 \begin{figure*}
	\centering
	\begin{tabular}{ccc}
		\begin{overpic}[width=75mm]{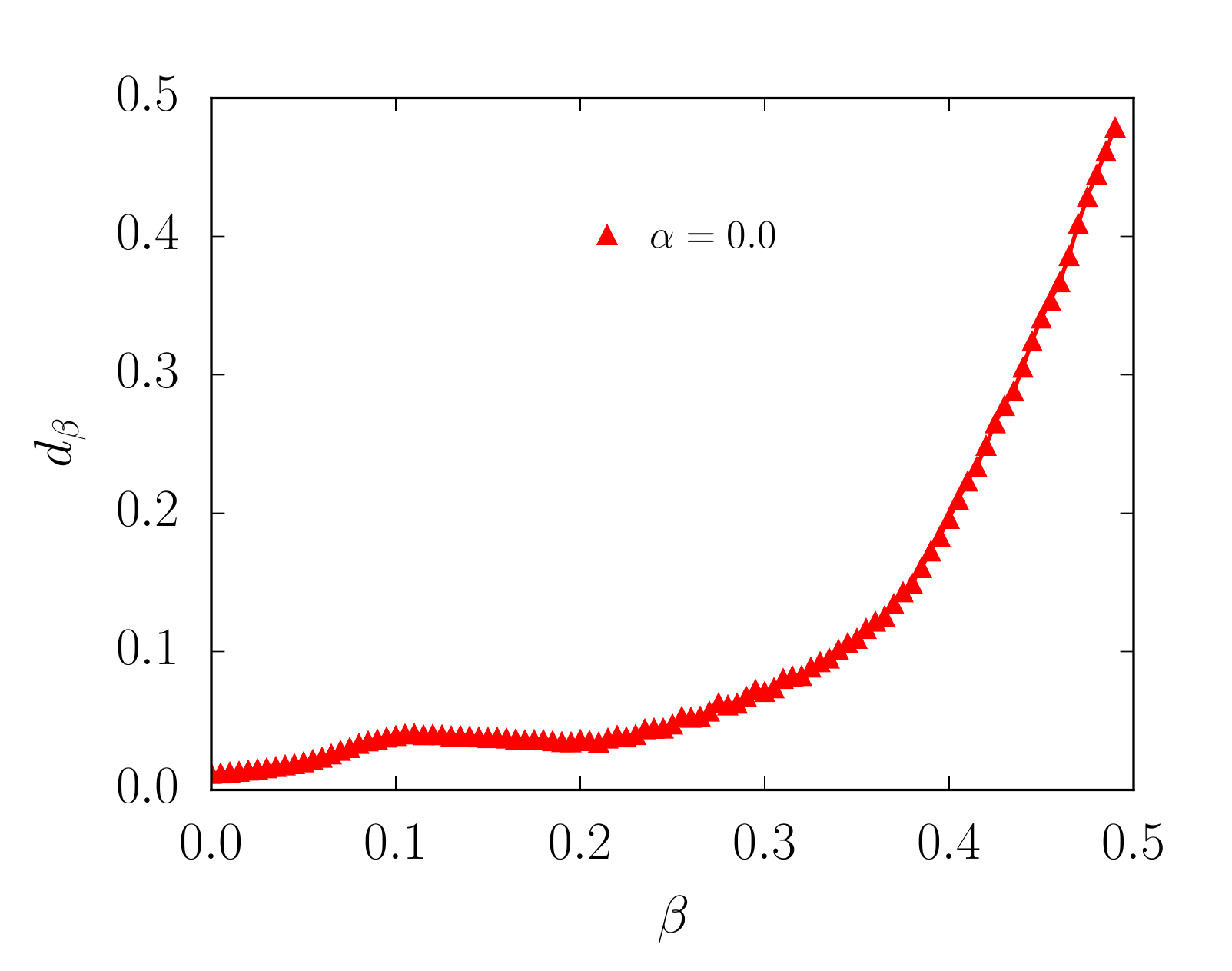}\put(-1,69){(a)}\end{overpic}&
		\begin{overpic}[width=75mm]{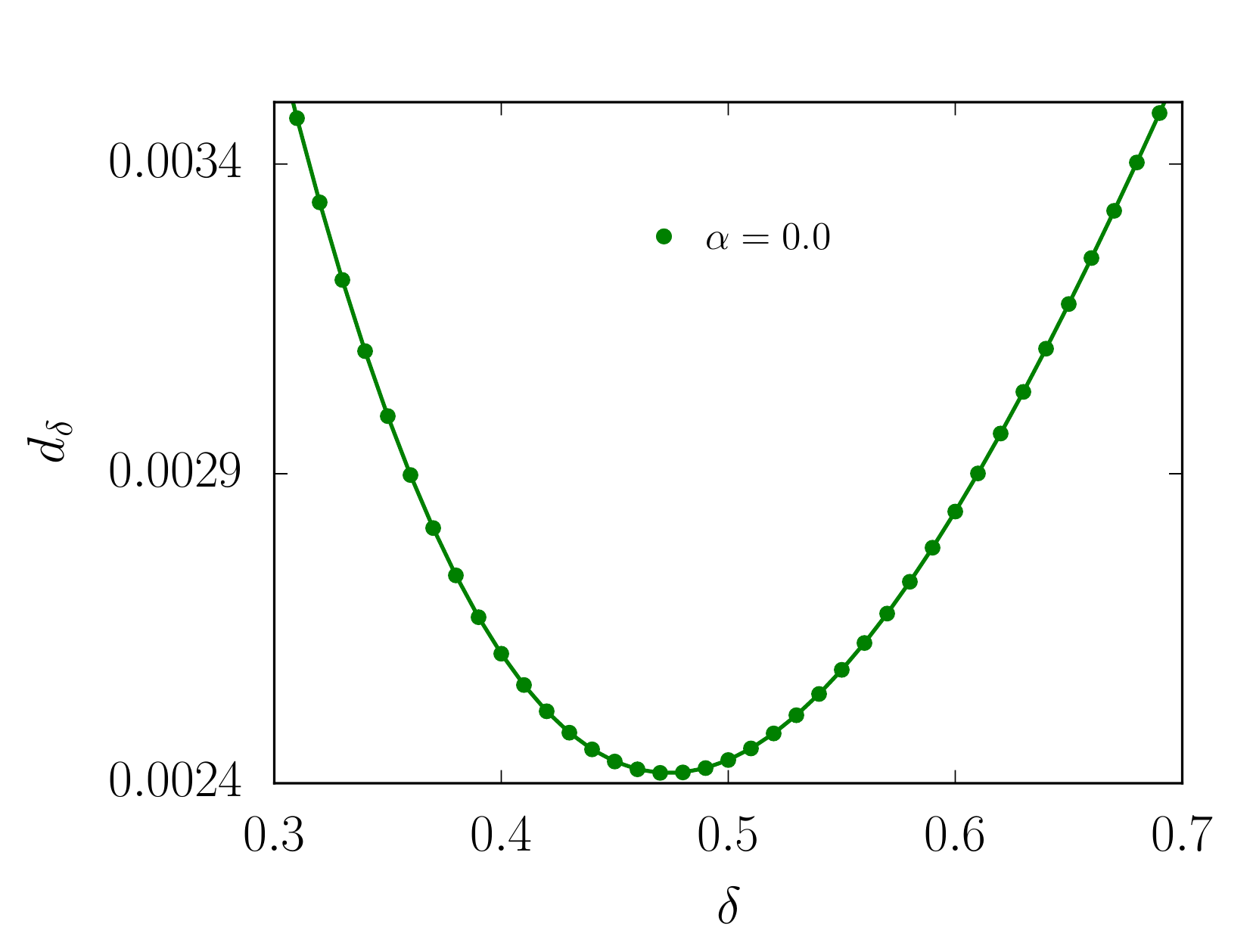}\put(-1,69){(b)}\end{overpic}
	\end{tabular}
     \caption{(a): Minimum distance between curves in Fig.~\ref{plotmm:fig}(a) as a function of $\beta$. (b): Cost function as function of $\delta$. Here $\alpha=0.0$.}
	\label{fig:ditance_exponent}
\end{figure*}}

\begin{thebibliography}{42}
\bibitem{note_th} 
Exact diagonalization is performed exploiting the fact that the model for $\alpha=0$ conserves the total-spin operator and one can restrict to the subspace with total spin $N/2$. For details see for instance~\cite{angelo_th}\BibitemShut {NoStop}
\end{thebibliography}
\end{document}